\documentclass{olplainarticle}

\title{Quantifying Flow separation for ellipse and von-Kármán Airfoil: A dataset of surface pressure and skin friction}

\author[1,2]{Christian Bak Winther}
\author[1]{Peter Ammundsen}
\author[1]{Fynn Jerome Aschmoneit\thanks{Corresponding author: fynnja@math.aau.dk}}
\affil[1]{Department of Mathematical Sciences, Aalborg University, Denmark}
\affil[2]{Propeller \& Aft Ship Department, Everllence SE, Denmark}

\usepackage{graphicx} 
\usepackage{amsmath}
\usepackage{pdfpages}
\usepackage{pdflscape}
\usepackage{subcaption}
\usepackage{import, graphicx, xcolor, calc}
\usepackage{hyperref}
\usepackage{float}
\usepackage{svg}
\usepackage{booktabs}

\newcommand{\mywidthTWO}{0.32\textwidth}
\newcommand{\mywidthdres}{0.475\textwidth}

\begin{document}

\maketitle

\section{Introduction}

The finite volume method is widely used to simulate flow around complex geometries. 
In many engineering applications, however, less computationally expensive models are preferred. 
In order to solve specific problems, potential flow models have been extended with concepts such as the Kutta-condition and viscous-inviscid coupling. 
Flow separation in particular is still, however, challenging to capture in a potential flow model. 


While viscous effects have been coupled to potential flow models, such as in the potential flow code XFOIL \cite{drela_viscous-inviscid_1987}, \cite{drela_xfoil_1989}, many assumptions are typically embedded in the implementation.  On blunt-bodies, such as ellipses, flow separation is expected to occur at moderate Reynolds numbers, at any angle of attack. 


For the development of such models, that may account for flow separation captured by potential flow, high quality reference data is needed for calibration and validation purposes. 
There is a limited availability of such high-fidelity, openly accessible datasets describing separated flow around simple geometries. While lift and drag coefficients can be found, skin-friction and separation points are typically not provided. 
The ellipse is of particular interest in this context, as it exhibits flow separation at all angles of attack at moderate Reynolds numbers. Additionally, the shape, and its derivatives are easily described for an ellipse.  
The dataset presented here provides numerically resolved flow fields around an ellipse and a Kármán-Trefftz airfoil. The results presented here are intended to support the development, calibration, and assessment of extended potential flow models.

This dataset contains local pressure- and skin friction coefficients, obtained through simulation in OpenFOAM. Stagnation points and separation points are determined through postprocessing.

\section{Geometry and numerical setup}

The simulations are performed using OpenFOAM v2506, with the simplefoam solver. Turbulence is modeled with the RANS method using the Menter $k\omega SST$ model \cite{menter_two-equation_1994} turbulence model. The model is chosen for its applicability on general flows, as well as native support of \textit{Low-Re} wall treatment.  

The $k\omega SST$ model assumes a completely turbulent, or \textit{tripped} boundary layer, and thus does not model laminar-turbulent transition. This may affect the results in two ways. Firstly, laminar boundary-layers are more prone to separation in adverse pressure gradients. Thus, $k\omega SST$ may fail to predict laminar separation, were it to occur.
In the present work, however, the Reynolds number is sufficiently large that separation is always expected to occur in the turbulent region.
The second effect comes in the form of boundary-layer thickness. 
Under free transition the existence of a the laminar region preceding the turbulent transition, results in a thinner overall boundary-layer. Thinner boundary-layers tend to withstand larger adverse pressure gradients before separation. Thus, with ideal transition modeling, one might see separation points move slightly downstream. 
There exists transition models for the $k\omega SST$ model, such as $\gamma-Re_{\theta \: t} $ \cite{langtry_correlation-based_2009}. This model uses empirical correlations to predict the onset of turbulence. For the sake of reproducibility it was elected not to use such models. 

Low-Re wall treatment is used, meaning the boundary layer is resolved including the viscous sub-layer. To satisfy requirements for low-Re wall treatment the domain is meshed to achieve a dimensions-less wall distance to the first cell-center, $y^+$, around unity. Separate meshes are created for each Reynolds number to comply with this $y^+$ requirement. The same mesh is, however, used across all angles of attack. Cell-to-cell expansion ratio normal to the wall is smaller than 1.2 globally. The mesh is 2 dimensional, consisting of a single cell in the z-direction.


\subsection{Geometry}

The major axis of the ellipse is used for the chord length, $c$. The ellipse posses a semi-major- and semi-minor axis of 0.5 and 0.125 respectively. The airfoil has a thickness-ratio $t/c = 0.1457$, it is symmetrical and the trailing edge angle is $0^\circ$. The Kárman-Trefftz airfoil has been constructed according to the method outlined in \cite{milne-thomson_theoretical_1973}. The ellipse and the von Kármán airfoil can be seen in figures \ref{fig:ellispe_form} and \ref{fig:vk_foil_form} respectively.

\begin{minipage}{\mywidthdres}
    \centering
    \includegraphics[width=\linewidth]{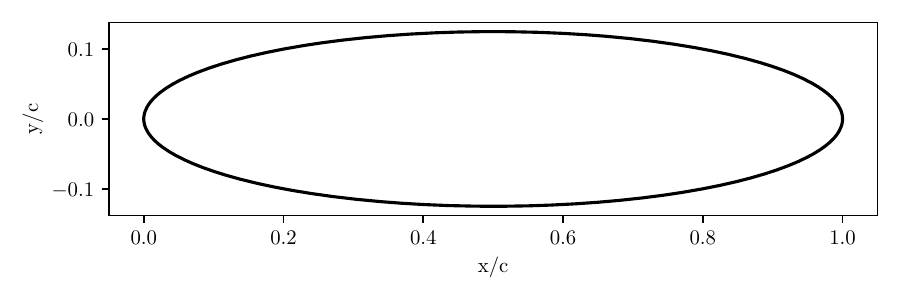}
    \captionof{figure}{Ellipse geometry}
    \label{fig:ellispe_form}
\end{minipage}
\hfill
\begin{minipage}{\mywidthdres}
    \centering
    \includegraphics[width=\linewidth]{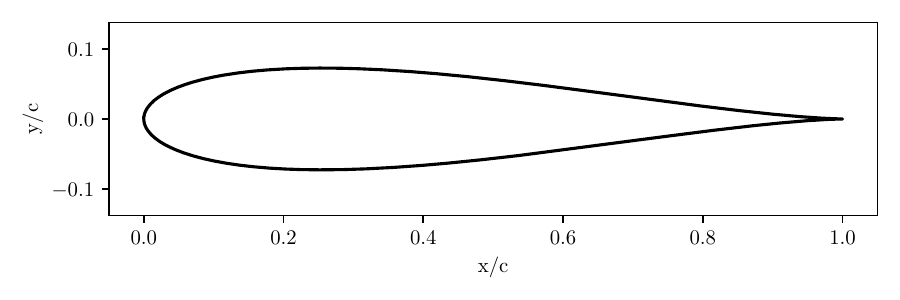}
    \captionof{figure}{von Kármán airfoil}
    \label{fig:vk_foil_form}
\end{minipage}



\subsection{Turbulence model and boundary conditions}

The desired ambient flow, the flow close to the model, is found from recommendations by Menter \cite{menter_two-equation_1994} and Spalart et al \cite{spalart_effective_2007}. 
From these recommendations, the specific dissipation rate is set to: 
\begin{equation} \label{eq:omega}
    \omega = \frac{5 ||\mathbf{U}||}{c}   
\end{equation}
The turbulent viscosity ratio is found as Reynolds dependent: 
\begin{equation}\label{eq:nutOverNu}
 \frac{\nu_t}{\nu} = 2\cdot 10^{-7}Re_c   
\end{equation}
The turbulent viscosity ratio is used here, as it is believed to be more generally applicable compared to the more common velocity-dependent $k$-recommendations. Turbulent kinetic energy is found from: 
\begin{equation} \label{eq:tke}
    k = \nu_t\omega
\end{equation}

To circumvent turbulent decay from the inlet to the geometry, $k_\infty$ and $\omega_\infty$ are set as both inlet conditions and as a minimum value in the simulation. 
This way the same ambient flow is present regardless of distance to inlet or mesh resolution. 
Since both $k$ and $\omega$ will increase in proximity to walls, imposing a lower limit will not influence the results negatively. This minimum value method is in line with the recommendations by Spalart et al \cite{spalart_effective_2007}. 

The angle of attack is changed by adjusting the inlet velocity vector. Inlet celerity is kept constant across all simulations at $10 \:m/s$. Reynolds number is adjusted by modifying the kinematic viscosity. This way, given equations \ref{eq:omega}, \ref{eq:nutOverNu} and \ref{eq:tke} the ambient specific dissipation rate, ambient turbulent kinematic viscosity and ambient turbulent kinetic energy are all constant across Reynolds numbers, respectively taking on the values $\omega_\infty = 50 \: [s^{-1}]$, $\nu_{t \: \infty} = 2\cdot 10^{-6} \: [m^2/s]$ and $k_\infty = 2.5\cdot 10^{-4} \:[m^2/s^2]$. 


    



\section{Mesh convergence}
The simulation results are verified through a mesh convergence and turbulence sensitivity analysis. 
A single representative case is selected for the mesh convergence study. The chosen case is the ellipse at $Re=10^6$, $\alpha = 10^\circ$. For each mesh the total cell count, $N_{cells}$, the approximate first layer thickness, $\delta_1$, and the cell-to-cell expansion ratio normal to the surface, $\chi$, are given in table \ref{tab:mesh_convergence_res}. 

\begin{table}[H]
    \centering
    \begin{tabular}{l|ccc|cccc|c}
    \toprule
    Mesh &  $N_{cells}$ & $\delta_1$, $[\mu m]$ & $\chi$ &  $C_d$ & $C_l$ & ss sep. x/c & ps sep. x/c & $\varepsilon_{C_d}$\\
    \midrule
    Coarse          & 23800 & 55.6 & 1.12 & 0.03351 & 0.69695 &  0.86495 & 0.95744 & -\\
    Medium-coarse  & 36000   & 48.8 & 1.10 & 0.03308 & 0.70183 &  0.86537 & 0.95703 & 1.30\% \\
    Medium          & 48000  & 48.0 & 1.08 & 0.03276 & 0.70370 &  0.86532 & 0.95653 & 0.98\% \\
    Medium-fine     & 75400 & 40.6 & 1.07  & 0.03261 & 0.70612 &  0.86508 & 0.95596 & 0.46\% \\
    Fine            & 102000  & 34.8 & 1.06& 0.03257 & 0.70595 &  0.86473 & 0.95565 & 0.12\% \\
    \bottomrule
    \end{tabular}
    \caption{Simulation results for different mesh resolutions}
    \label{tab:mesh_convergence_res}
\end{table}

\begin{figure}[htbp]
    \centering
    \begin{subfigure}{0.48\textwidth}
        \centering
        \includegraphics[width=\linewidth]{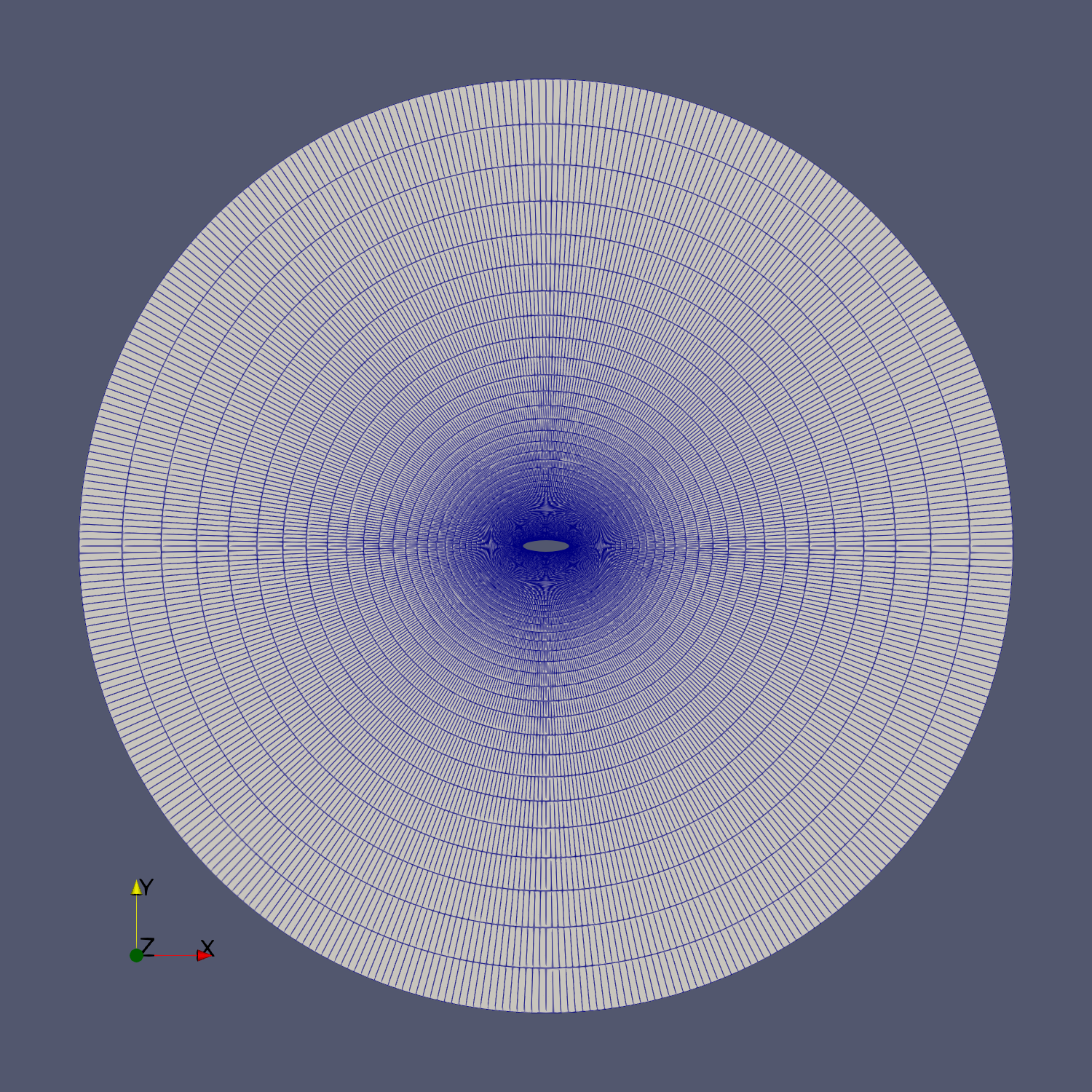}
        \caption{Ellipse meshed domain}
        \label{fig:img1}
    \end{subfigure}
    \hfill
    \begin{subfigure}{0.48\textwidth}
        \centering
        \includegraphics[width=\linewidth]{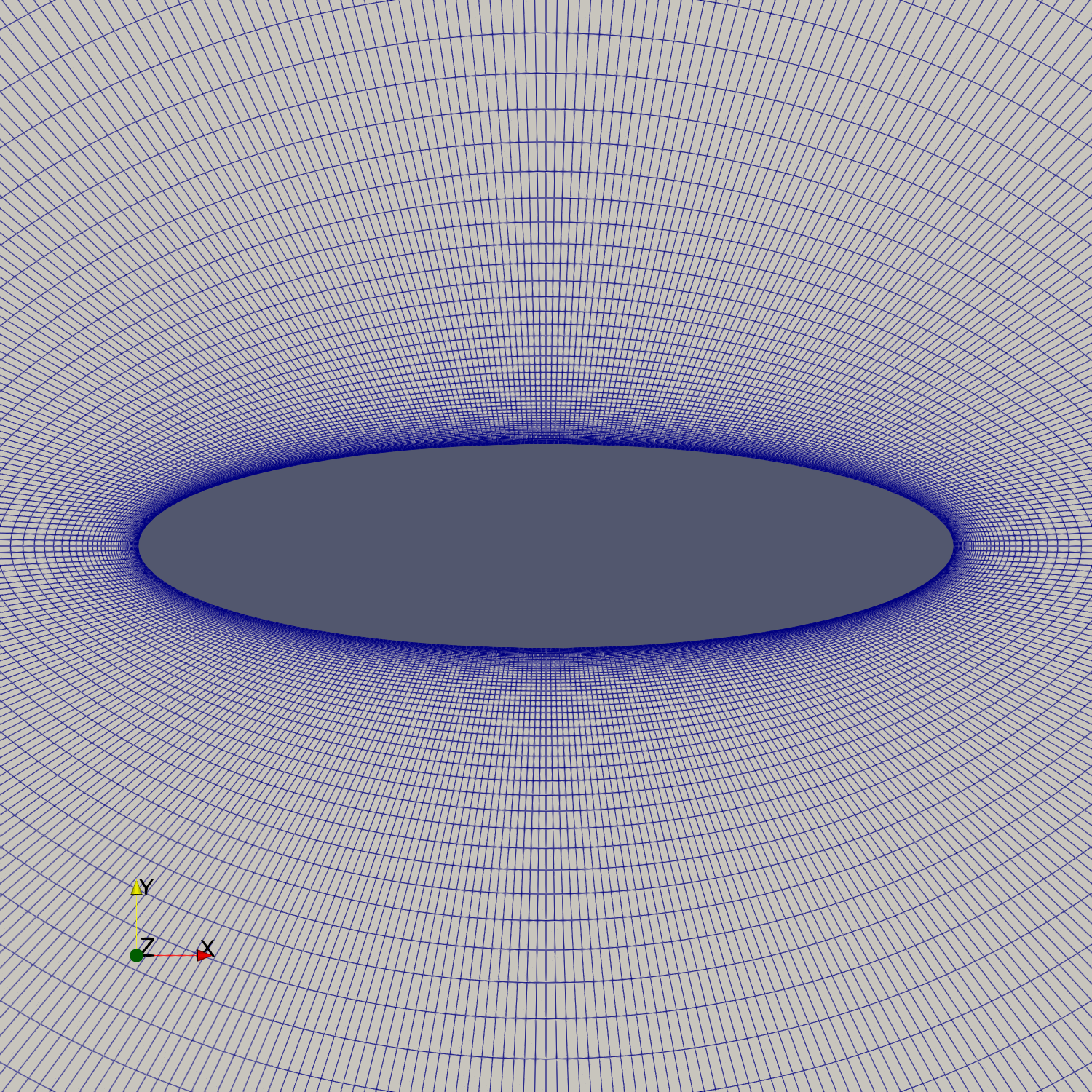}
        \caption{Closeup of ellipse mesh}
        \label{fig:img2}
    \end{subfigure}
    \caption{Ellipse mesh}
    \label{fig:EL_two_images}
\end{figure}
\begin{figure}[htbp]
    \centering
    \begin{subfigure}{0.48\textwidth}
        \centering
        \includegraphics[width=\linewidth]{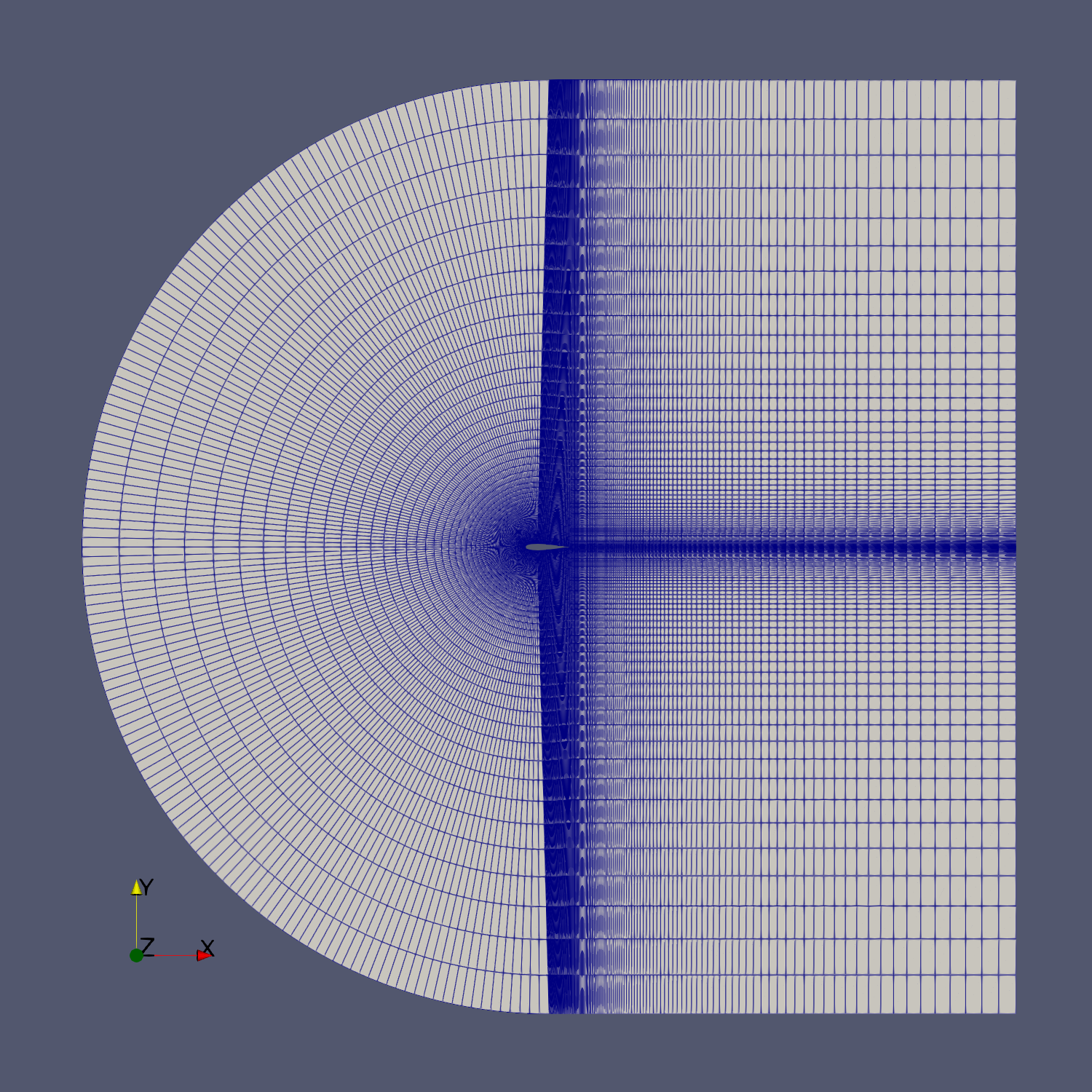}
        \caption{Airfoil meshed domain}
        \label{fig:img1}
    \end{subfigure}
    \hfill
    \begin{subfigure}{0.48\textwidth}
        \centering
        \includegraphics[width=\linewidth]{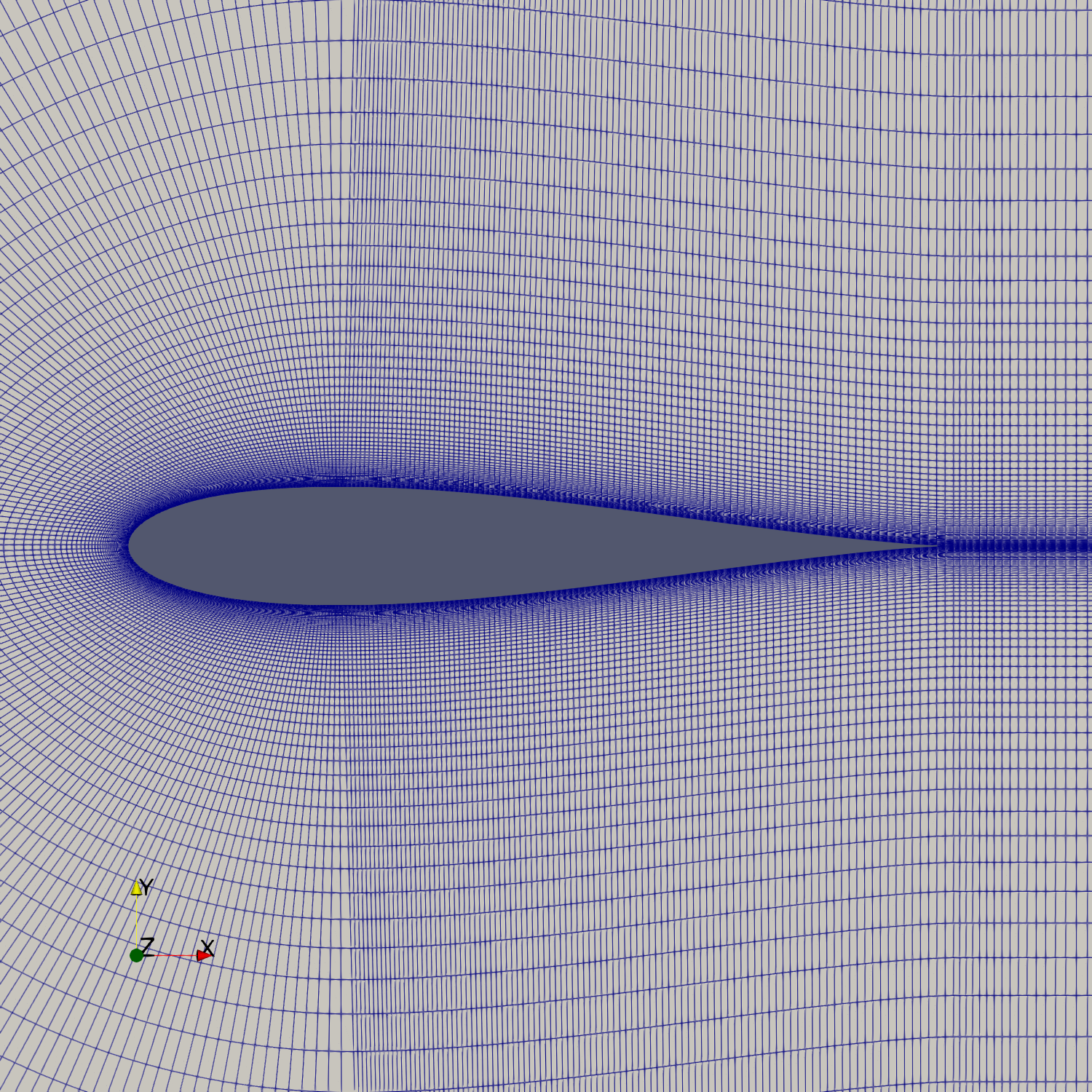}
        \caption{Closeup of airfoil mesh}
        \label{fig:img2}
    \end{subfigure}
    \caption{Airfoil mesh}
    \label{fig:VK_two_images}
\end{figure}

The following target variables are considered during the mesh sensitivity study: Drag coefficient $C_d$, Lift coefficient $C_l$, suction side- and pressure side separation points.

\begin{equation}\label{eq:qinf}
\begin{aligned}
     q_\infty = \frac{1}{2} \rho \:||\mathbf{U_\infty}||^2,
\end{aligned}
\qquad
\begin{aligned}
    C_d = \frac{D}{A q_\infty},
\end{aligned}
\qquad
\begin{aligned}
    C_l = \frac{ L }{A q_\infty}
\end{aligned}
\end{equation}
Lift and Drag coefficients are found according to equation \ref{eq:qinf}, where $q_\infty$ is the ambient dynamic pressure, $D$ and $L$ are total drag and lift forces respectively and $A$ is the planform area of the wing section, found as the chord length times the width of the mesh. 
Separation points on the ellipse are given in terms of the position along the chord $x/c$ "ss" and "ps" will be used to refer to the suction side and pressure side respectively. 
The drag coefficient is the most sensitive wrt. mesh resolution.
Its relative difference to its previous mesh resolution is given by $\varepsilon_{C_d}$, as shown in table \ref{tab:mesh_convergence_res}. 
For the medium mesh this error is less than 1\% and thus deemed sufficient for this study. The medium mesh will therefore be used for the remainder of this study. 
For the $Re=10^7$ mesh, cell-to-cell expansion ratio is adjusted to achieve the same $y^+$. Images of the medium mesh on the ellipse and airfoils can be seen in figures \ref{fig:EL_two_images} and \ref{fig:VK_two_images}, respectively.


\section{Turbulence sensitivity}
Selecting the appropriate ambient turbulence level in RANS simulations can be a non-trivial task. 
Generally, the turbulent kinetic energy field should only contain the turbulent energy at a length-scale relevant to boundary-layer \cite{spalart_effective_2007}. For this turbulence sensitivity study the same case is considered as for the mesh convergence study. 
In external flows the ambient turbulence level should not be so high as to interfere with the outmost regions of the boundary-layer. Additionally, the turbulence should not be so low as to render the turbulence model inactive. Generally external turbulent flow should be insensitive to ambient conditions within a reasonable range \cite{spalart_effective_2007}.  

The ambient turbulence level chosen for this study follows the recommendations by Spalart et al. in their discussion on the $k\omega SST$ model. Here a sensitivity study is conducted, quantifying the effect of turbulence on the separation point and force coefficients. 
To assess if these objectives are met in at the chosen ambient conditions, a turbulence sensitivity study is performed. The turbulence viscosity ratio is increased, while the specific dissipation rate is kept constant. The tested turbulence levels and their effect on target variables are shown in table \ref{tab:turbulence_test}.
\begin{table}[H]
    \centering
    \begin{tabular}{@{}l|ccc|cccc@{}}
    \toprule
    Case & $(\nu_t/\nu)_\infty$ $[-]$   & $\nu_{t \: \infty}$ $\left[m^2/s\right]$ & $k_\infty$ $\left[m^2/s^2\right]$ & $C_d$ & $C_l$ & ss sep. x/c & ps sep. x/c\\ 
    \midrule
    Baseline &  $0.2$ & $2.0\cdot 10^{-6}$ & $2.5\cdot10^{-4}$ & 0.03276 & 0.70370 & 0.86532 & 0.95653\\
    2x $\nu_t$ &  $0.4$ & $4.0\cdot 10^{-6}$ & $5.0\cdot10^{-4}$ & 0.03275 & 0.70396 & 0.86540 & 0.95652\\
    4x $\nu_t$ &  $0.8$ & $8.0\cdot 10^{-6}$ & $1.0\cdot10^{-3}$ & 0.03276 & 0.70376 & 0.86540 & 0.95653\\
    8x $\nu_t$ &  $1.6$ & $1.6\cdot 10^{-5}$ & $2.0\cdot10^{-3}$ & 0.03277 & 0.70363 & 0.86546 & 0.95656\\
    \bottomrule
    \end{tabular}
    \caption{Turbulence sensitivity test cases and effect on selected variables.}
    \label{tab:turbulence_test}
\end{table}



The shown values are indeed insensitive to ambient turbulence levels. Based on this analysis, it is concluded that the baseline turbulence settings used are appropriate for this study.  

\section{Results}

Simulations results are presented in the following order: First distributions of local force coefficients, friction and pressure. 
These data are found in the Supplementary Material, in file "Results.zip". 
Then, integral values of these values are presented in tables. Finally stagnation points and separation points are reported. 

\subsection*{Pressure coefficient and skin friction coefficient}

Surface pressure and skin friction are normalized by the ambient dynamic pressure $q_\infty$. This gives the pressure coefficient $C_p$ and skin friction coefficient $c_f$ respectively. 

\begin{equation}
\begin{aligned}
    C_p = \frac{p_\infty-p}{q_\infty},
\end{aligned}
\qquad
\begin{aligned}
    C_f = \frac{\text{sign}(\tau_x)\:||\tau||}{q_\infty}   
\end{aligned}
\end{equation}

\begin{figure}[htbp]
    \begin{minipage}{\mywidthTWO}
      \includegraphics[width=\textwidth]{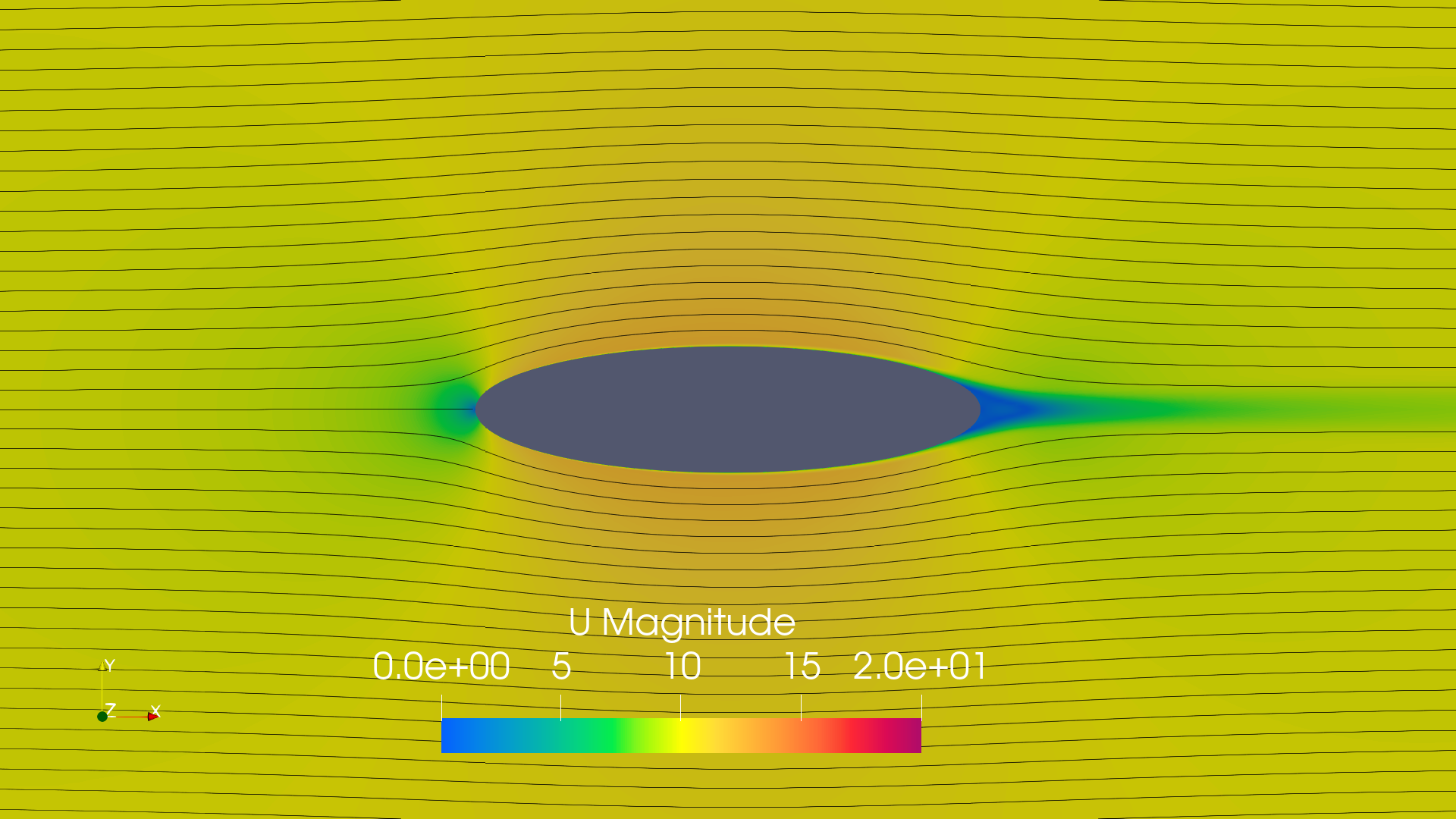}
      \end{minipage}\hfill
    \begin{minipage}{\mywidthTWO}
      \includegraphics[width=\textwidth]{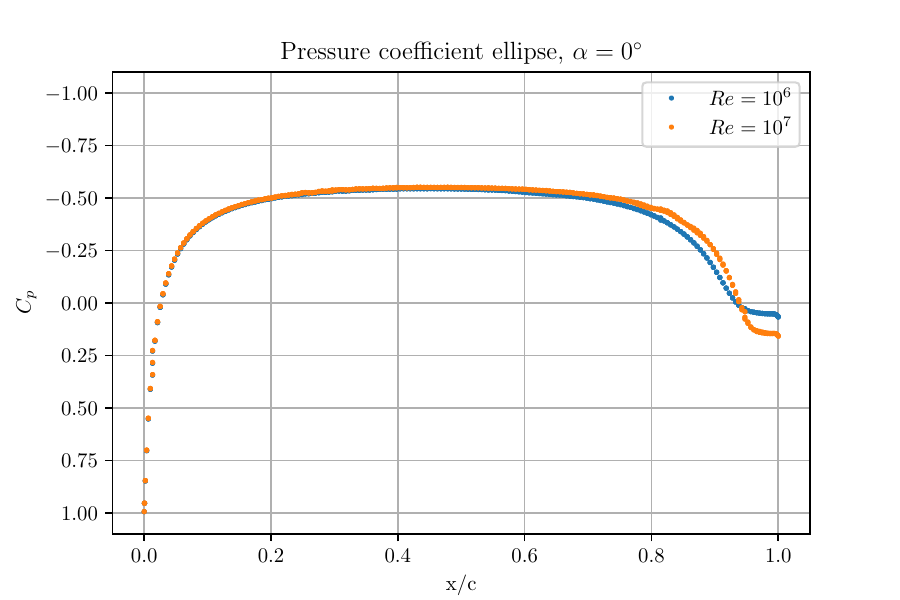}
      \end{minipage}\hfill
    \begin{minipage}{\mywidthTWO}
      \includegraphics[width=\textwidth]{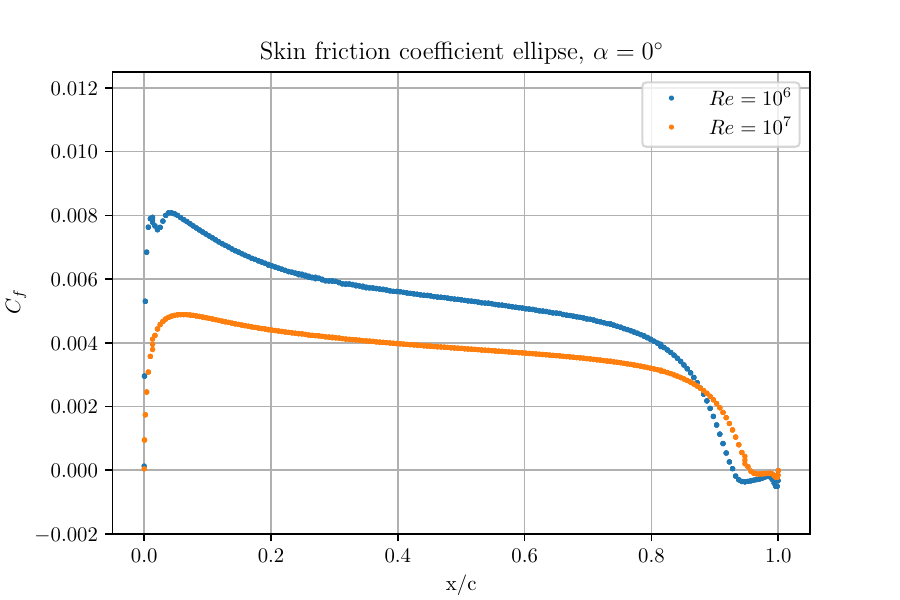}
      \end{minipage}
   \vspace{0.8em}
      \begin{minipage}{\mywidthTWO}
      \includegraphics[width=\textwidth]{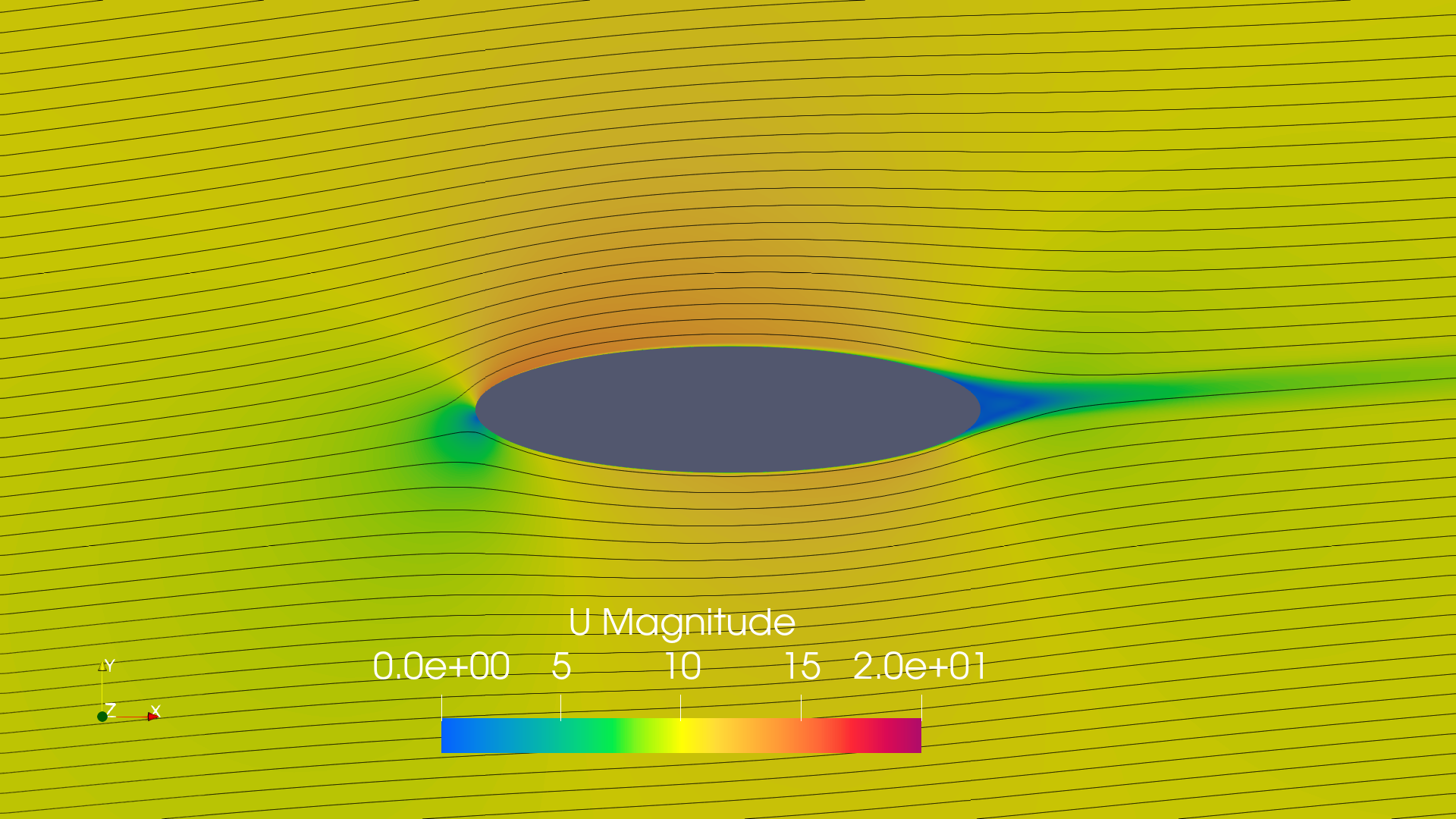}
      \end{minipage}\hfill
    \begin{minipage}{\mywidthTWO}
      \includegraphics[width=\textwidth]{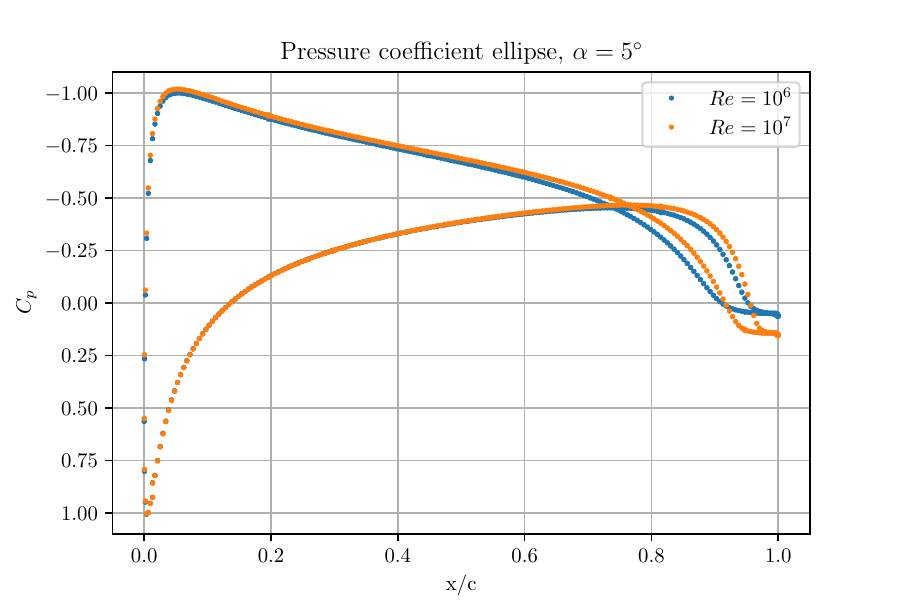}
      \end{minipage}\hfill
      \begin{minipage}{\mywidthTWO}
      \includegraphics[width=\textwidth]{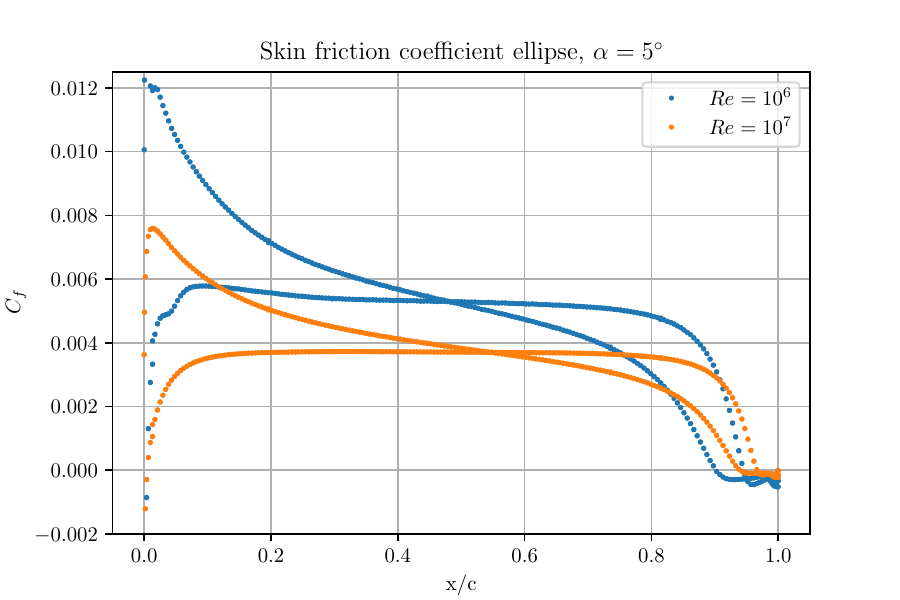}
      \end{minipage}
             \vspace{0.8em}
      \begin{minipage}{\mywidthTWO}
      \includegraphics[width=\textwidth]{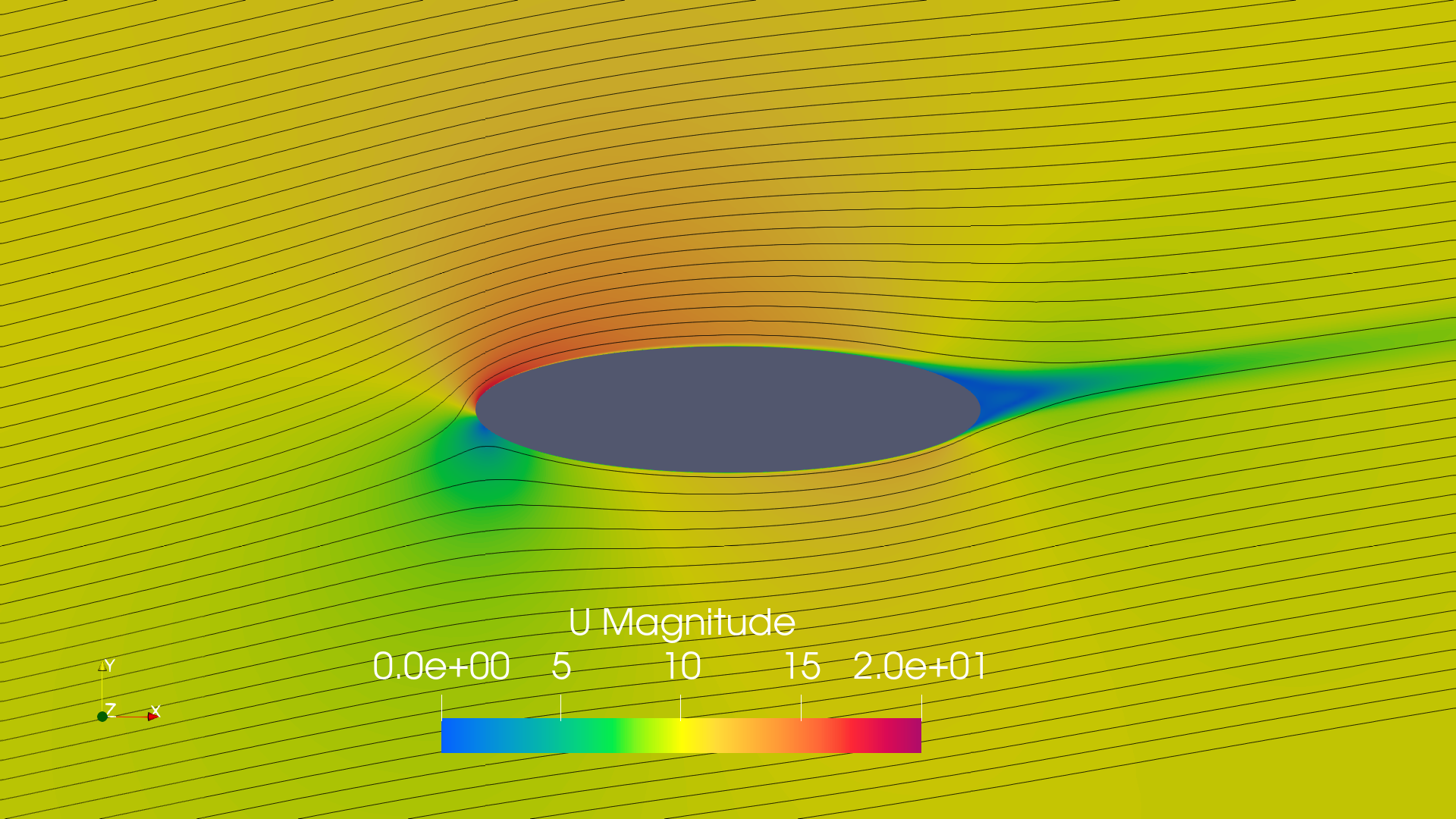}
      \end{minipage}\hfill
    \begin{minipage}{\mywidthTWO}
      \includegraphics[width=\textwidth]{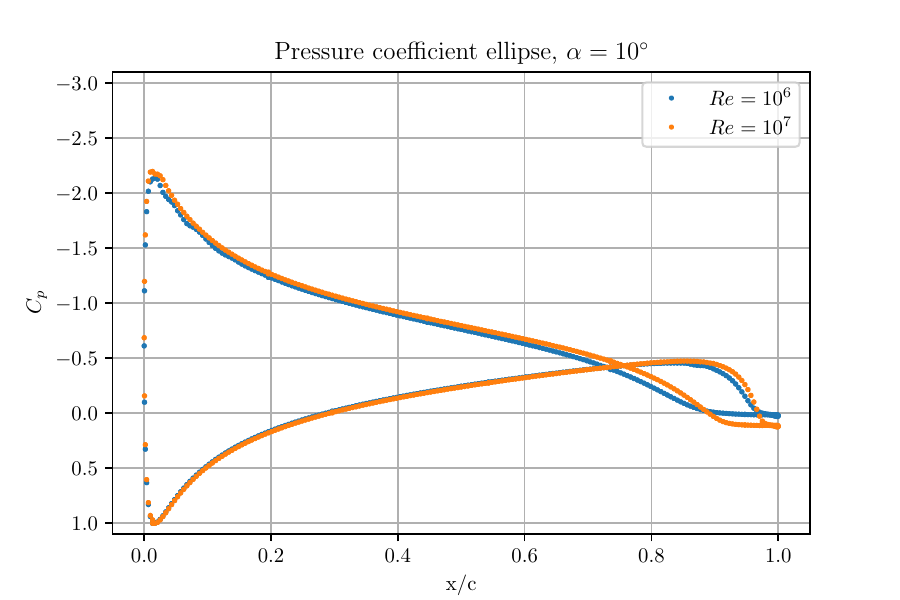}
      \end{minipage}\hfill
      \begin{minipage}{\mywidthTWO}
      \includegraphics[width=\textwidth]{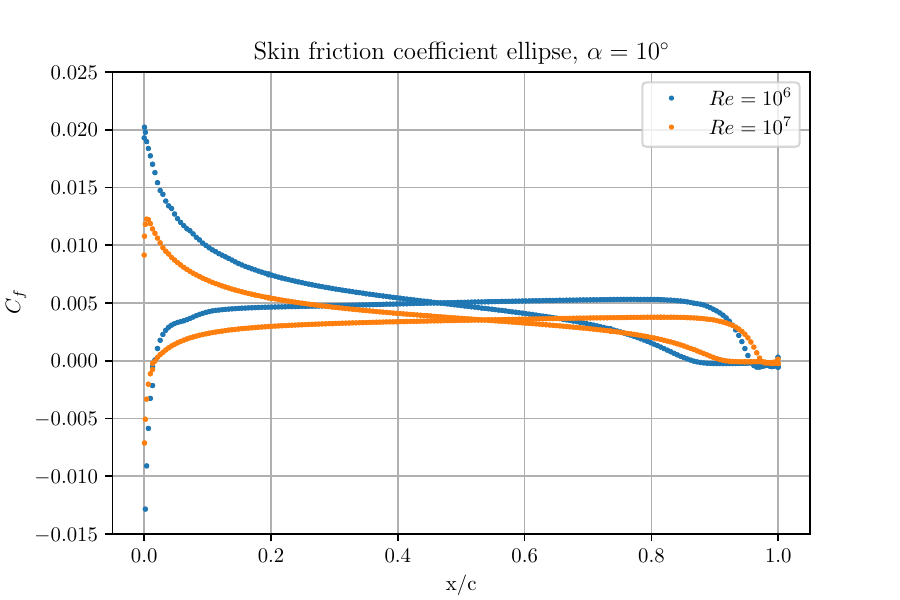}
      \end{minipage}
                         \vspace{0.8em}
      \begin{minipage}{\mywidthTWO}
      \includegraphics[width=\textwidth]{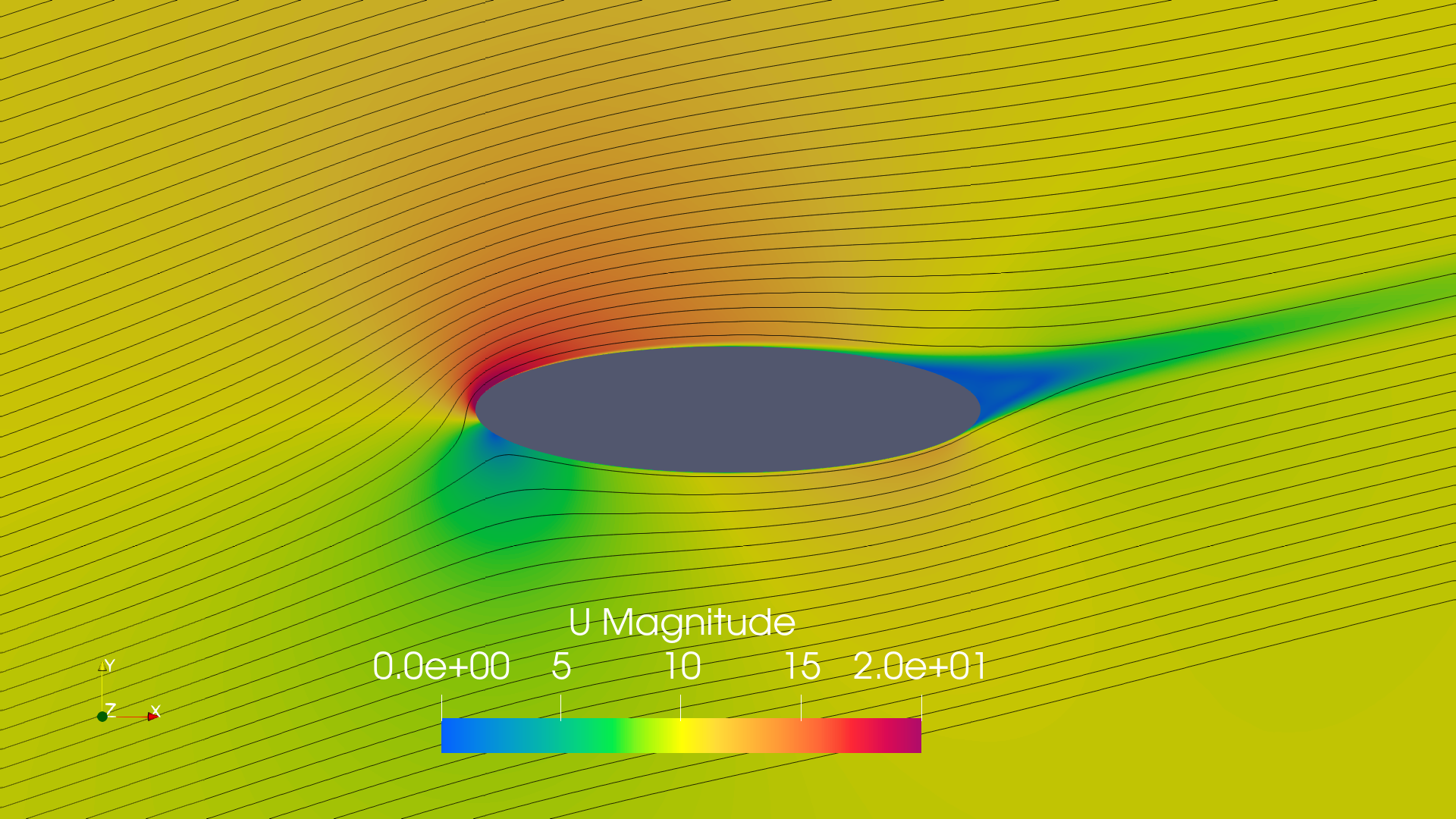}
      \end{minipage}\hfill
    \begin{minipage}{\mywidthTWO}
      \includegraphics[width=\textwidth]{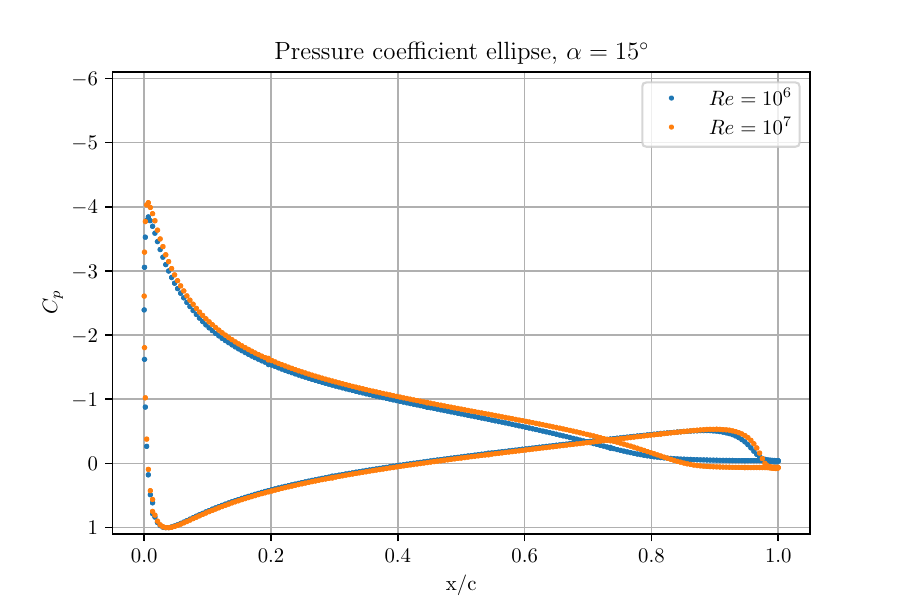}
      \end{minipage}\hfill
      \begin{minipage}{\mywidthTWO}
      \includegraphics[width=\textwidth]{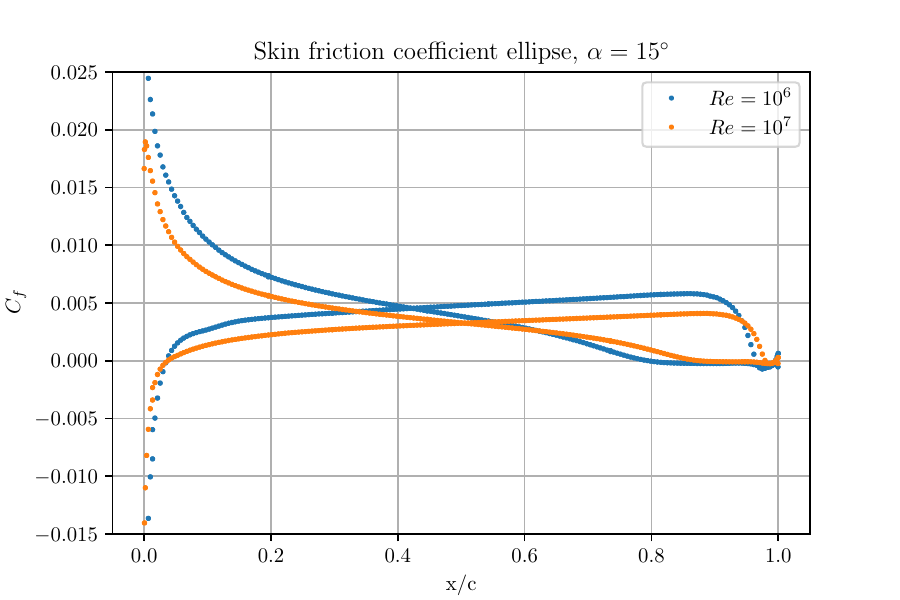}
      \end{minipage}
            \vspace{0.8em}
      \begin{minipage}{\mywidthTWO}
      \includegraphics[width=\textwidth]{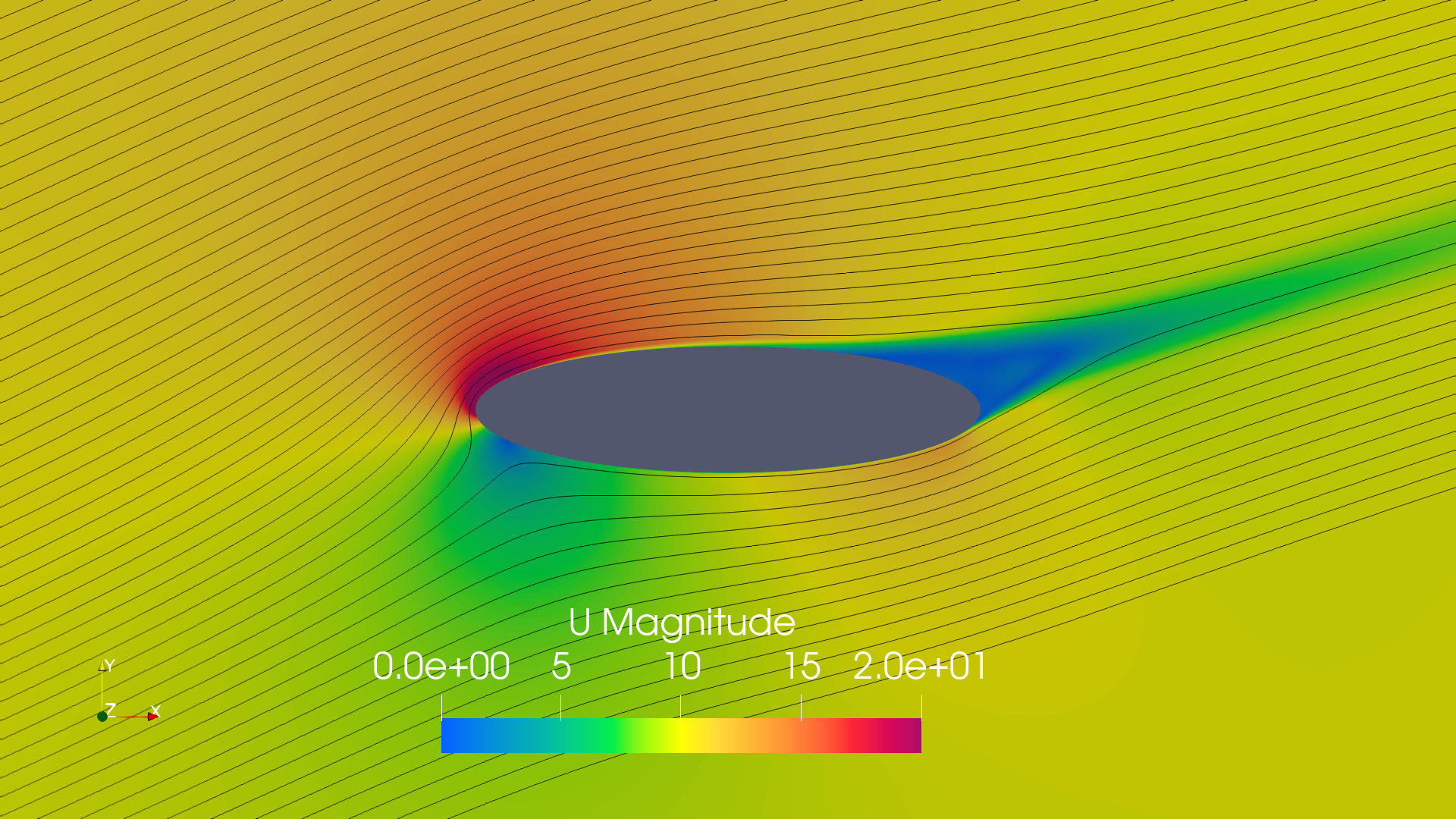}
      \end{minipage}\hfill
    \begin{minipage}{\mywidthTWO}
      \includegraphics[width=\textwidth]{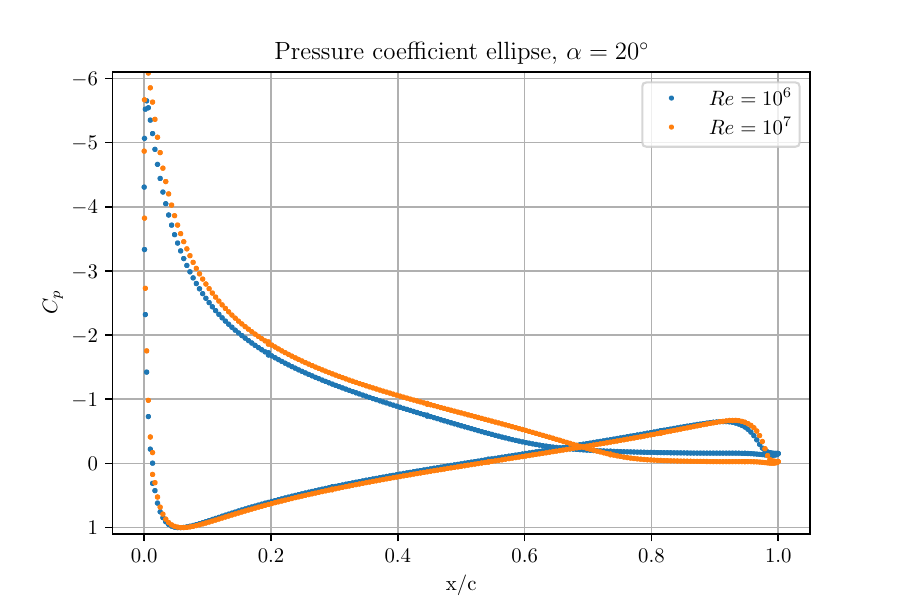}
      \end{minipage}\hfill
      \begin{minipage}{\mywidthTWO}
      \includegraphics[width=\textwidth]{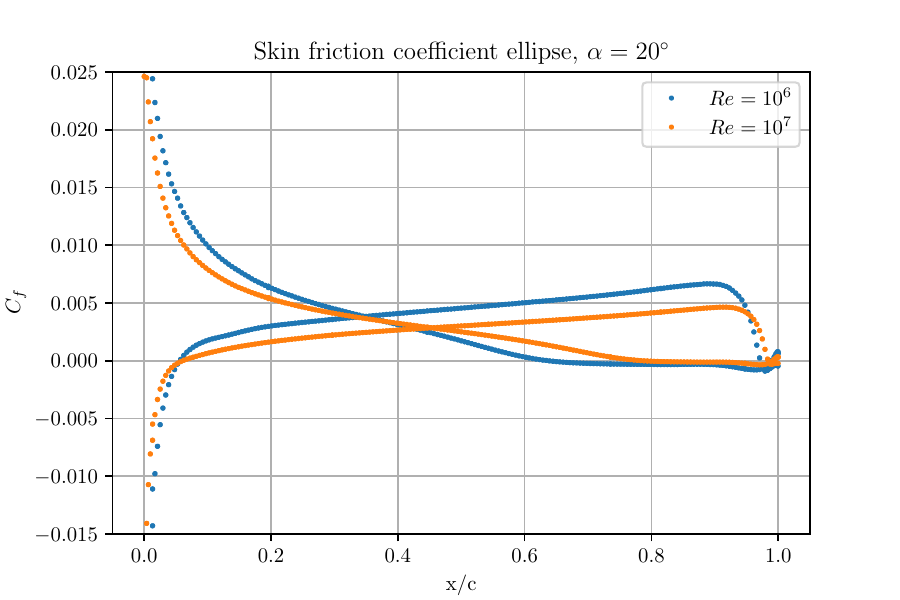}
      \end{minipage}
            \caption{Simulation results for the ellipse, Each row, from top to bottom; $\alpha = 0^\circ$, $\alpha = 5^\circ$, $\alpha = 10^\circ$, $\alpha = 15^\circ$ and $\alpha = 20^\circ$ . Each column, from left to right; streamlines at $Re = 10^7$, Pressure coefficient and Skin friction coefficient.}
                        \label{fig:ellispe_billeder}
\end{figure}

\begin{figure}[htbp]
  \centering
      \begin{minipage}{\mywidthTWO}
      \includegraphics[width=\textwidth]{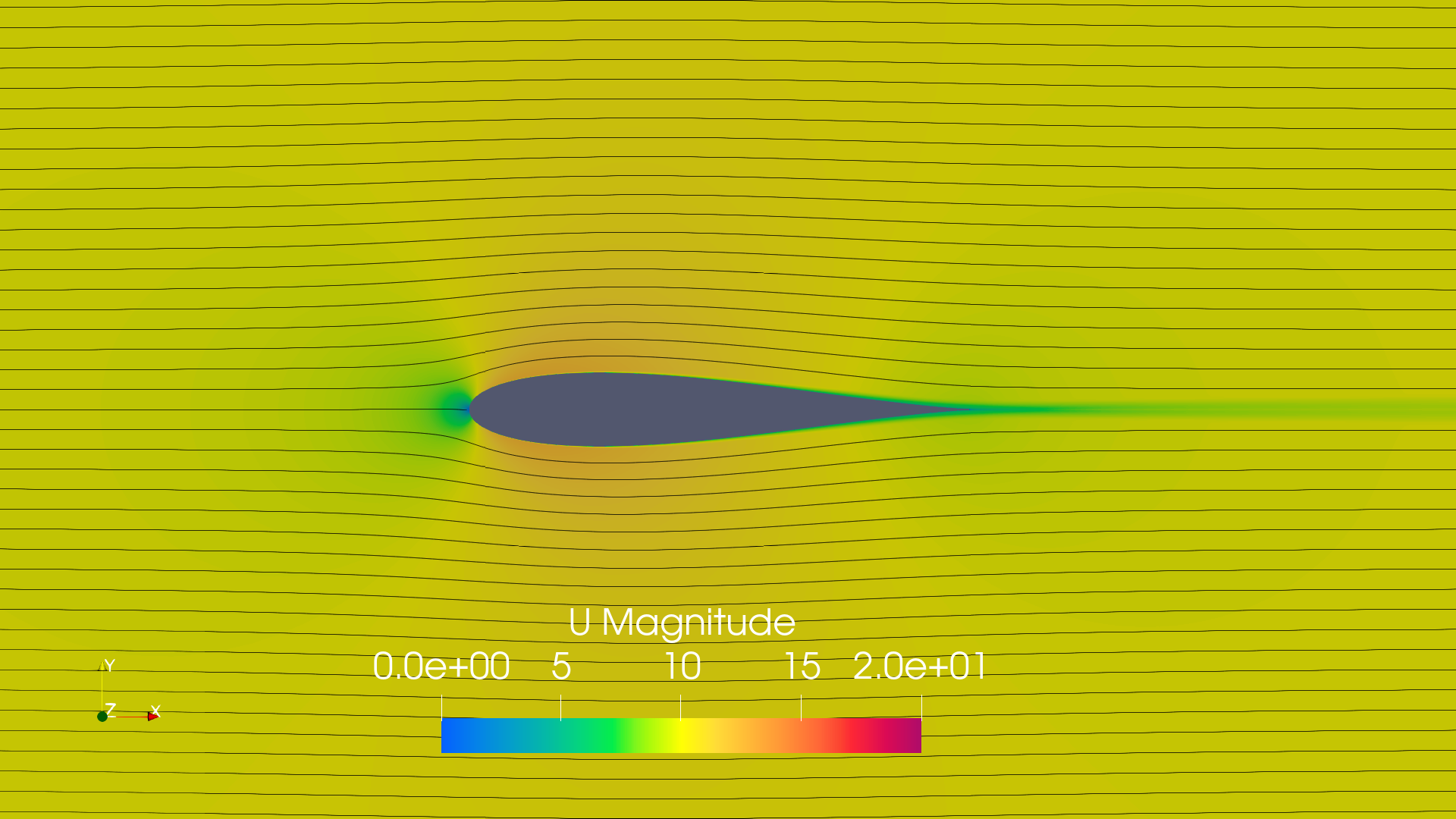}
      \end{minipage}\hfill
    \begin{minipage}{\mywidthTWO}
      \includegraphics[width=\textwidth]{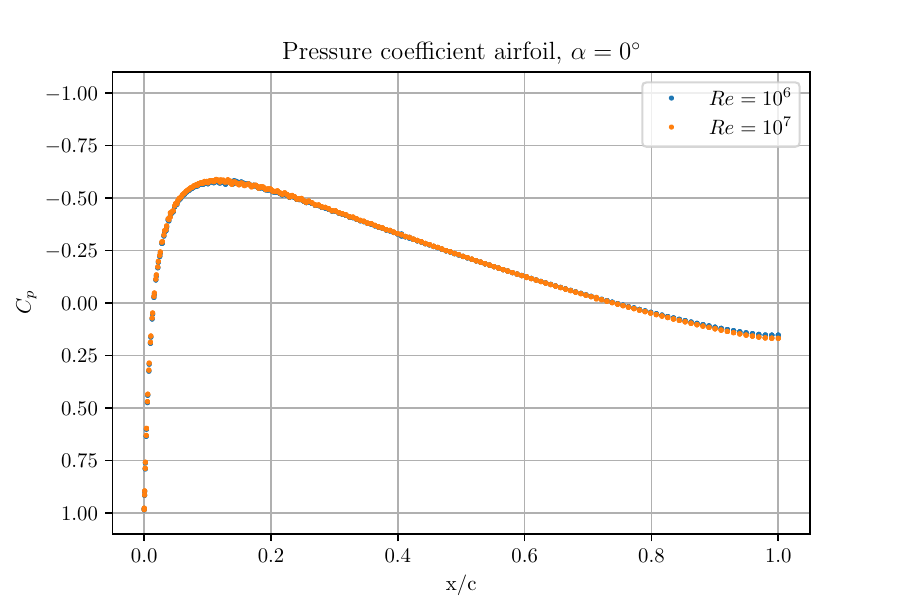}
      \end{minipage}\hfill
    \begin{minipage}{\mywidthTWO}
      \includegraphics[width=\textwidth]{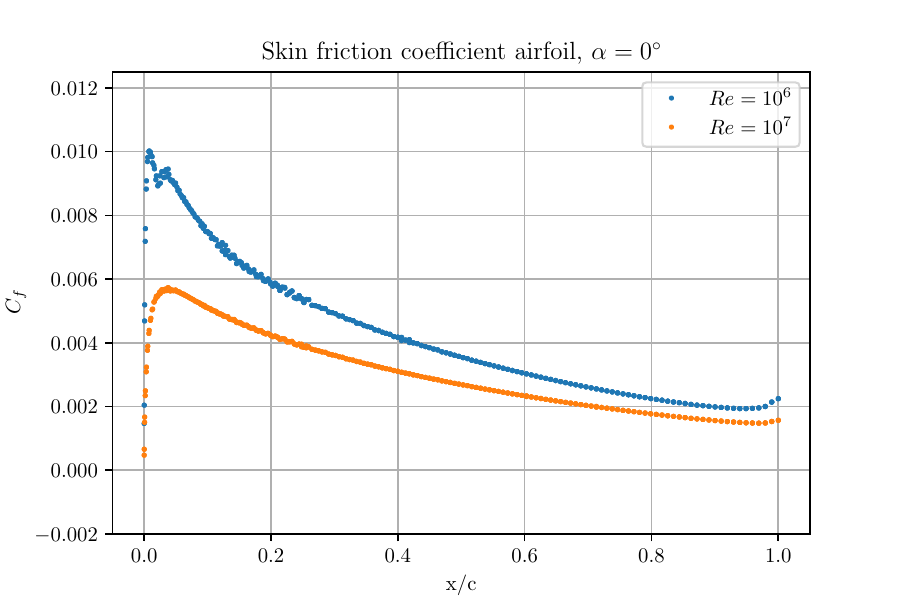}
      \end{minipage}
 \vspace{0.8em}
    \begin{minipage}{\mywidthTWO}
      \includegraphics[width=\textwidth]{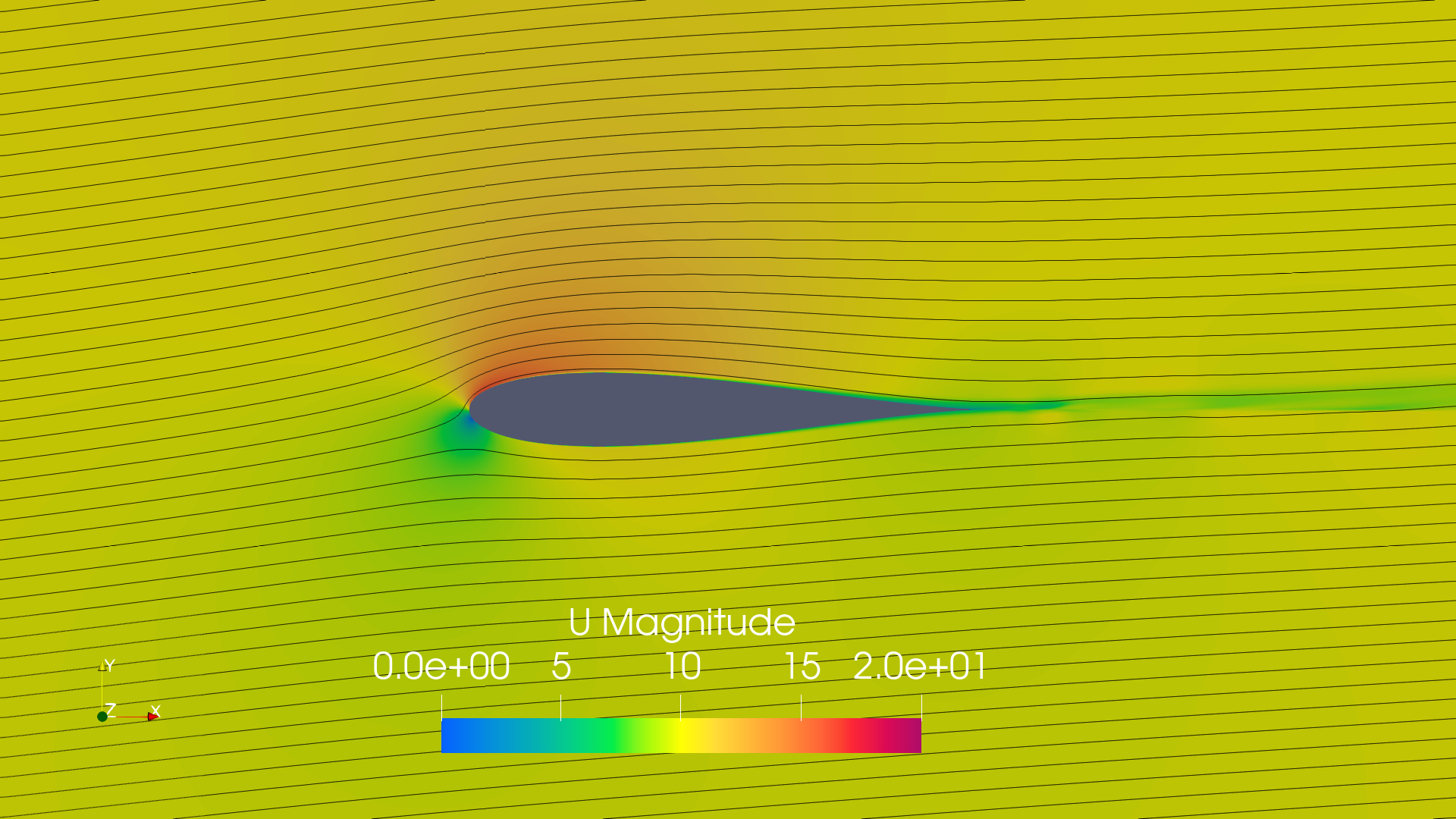}
      \end{minipage}\hfill
    \begin{minipage}{\mywidthTWO}
      \includegraphics[width=\textwidth]{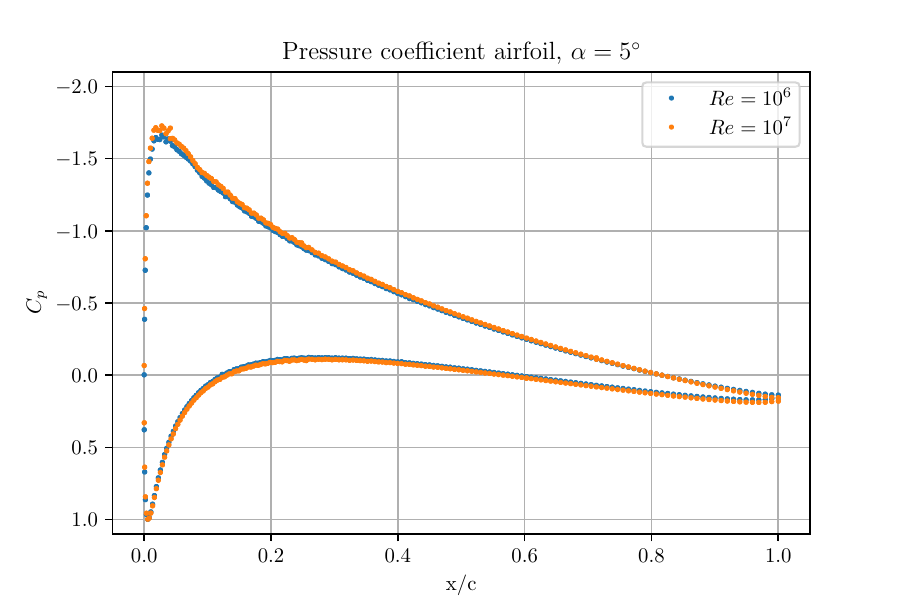}
      \end{minipage}\hfill
    \begin{minipage}{\mywidthTWO}
      \includegraphics[width=\textwidth]{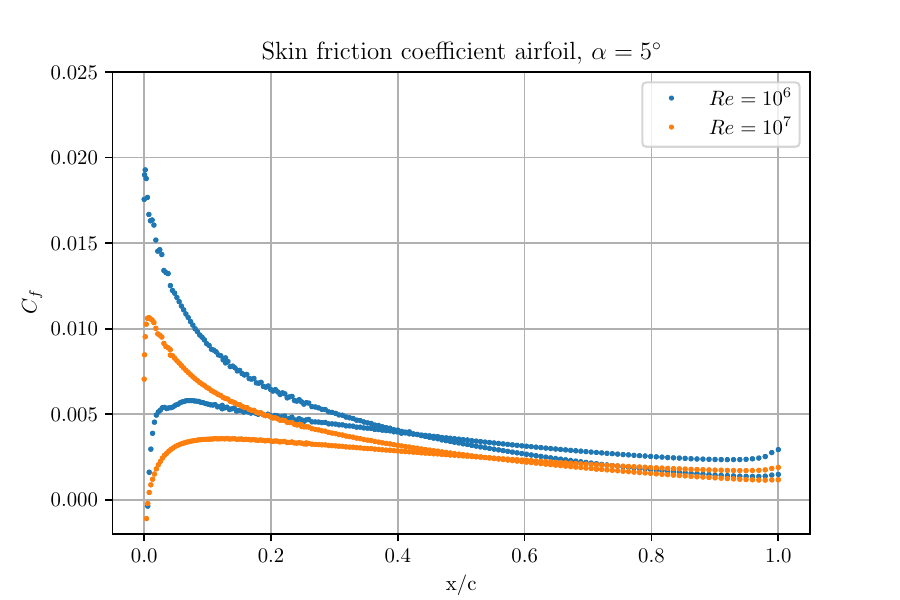}
      \end{minipage}
 \vspace{0.8em}
      \begin{minipage}{\mywidthTWO}
      \includegraphics[width=\textwidth]{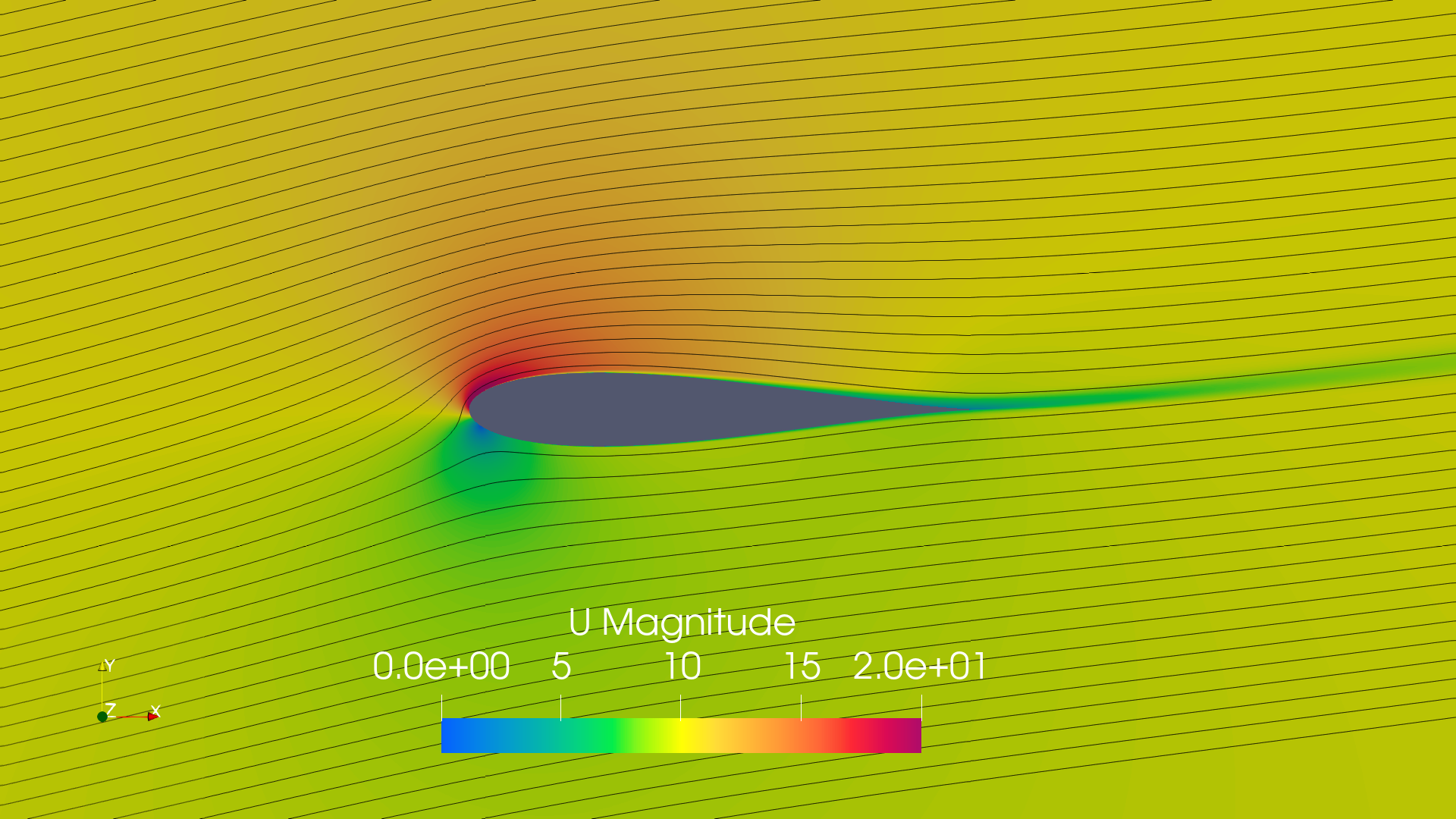}
      \end{minipage}\hfill
    \begin{minipage}{\mywidthTWO}
      \includegraphics[width=\textwidth]{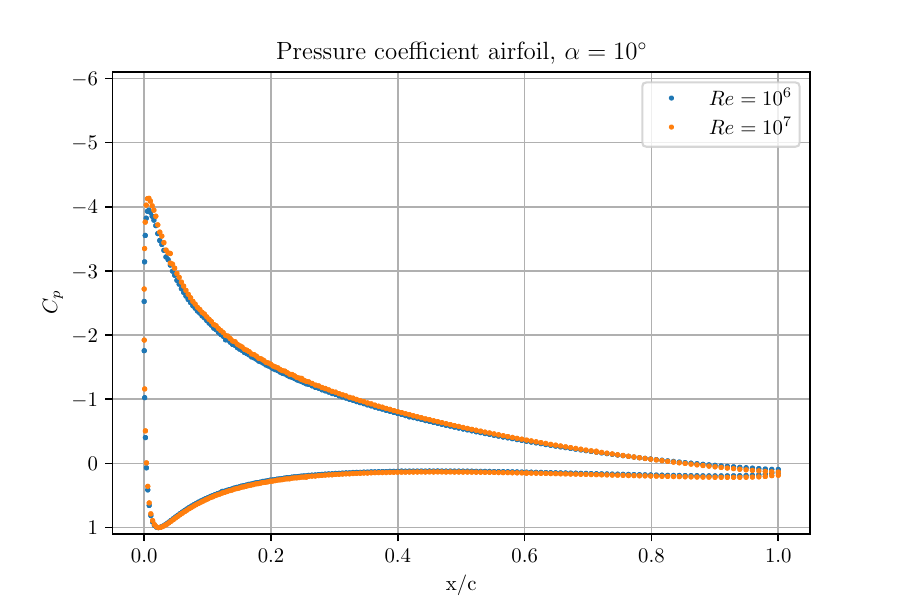}
      \end{minipage}\hfill
      \begin{minipage}{\mywidthTWO}
      \includegraphics[width=\textwidth]{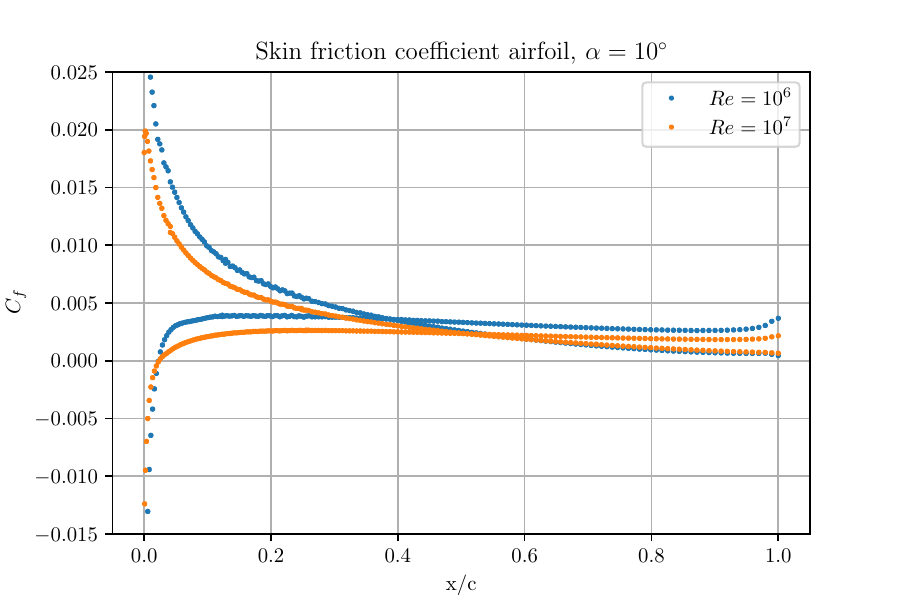}
      \end{minipage}
       \vspace{0.8em}
      \begin{minipage}{\mywidthTWO}
      \includegraphics[width=\textwidth]{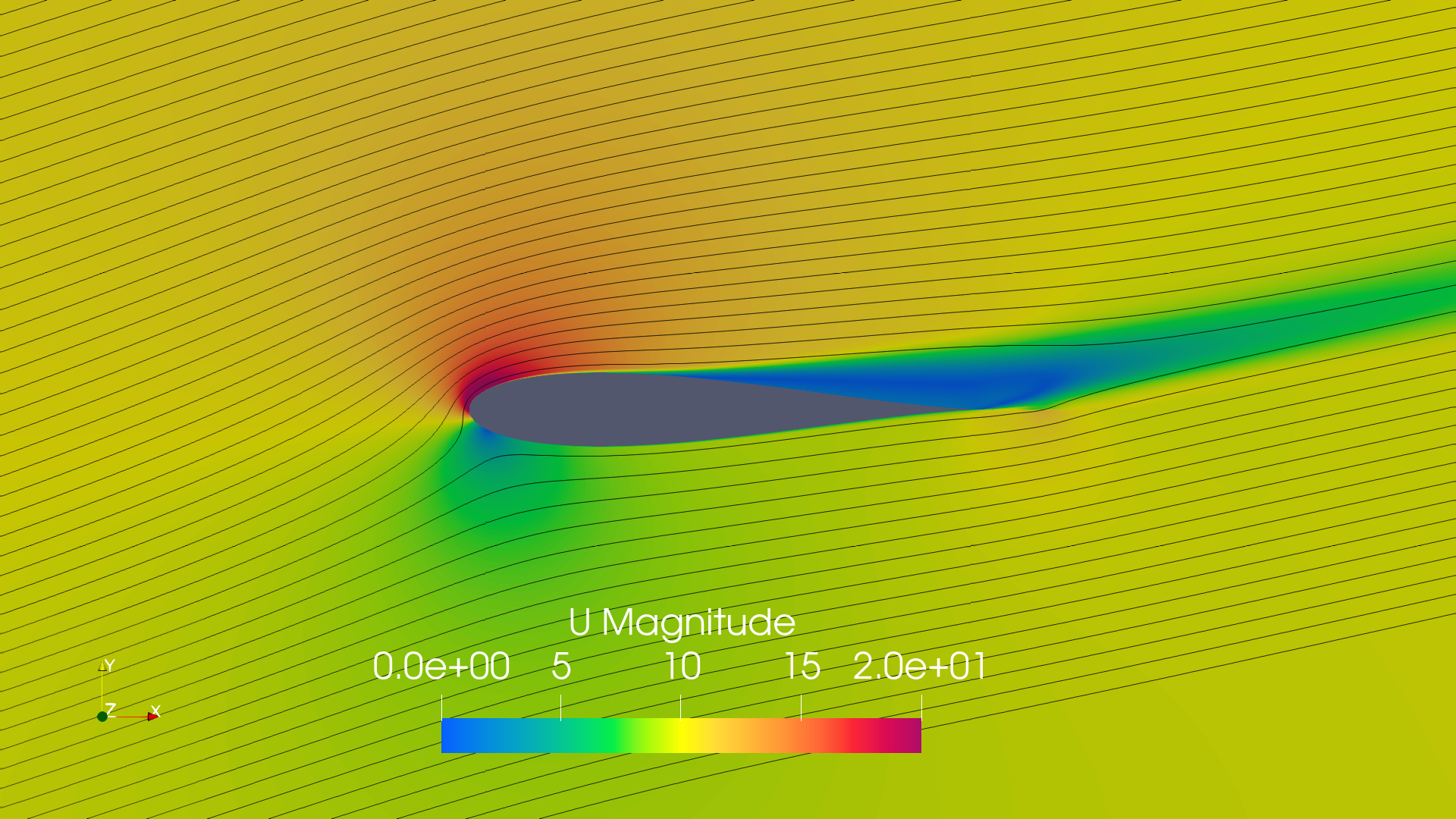}
      \end{minipage}\hfill
    \begin{minipage}{\mywidthTWO}
      \includegraphics[width=\textwidth]{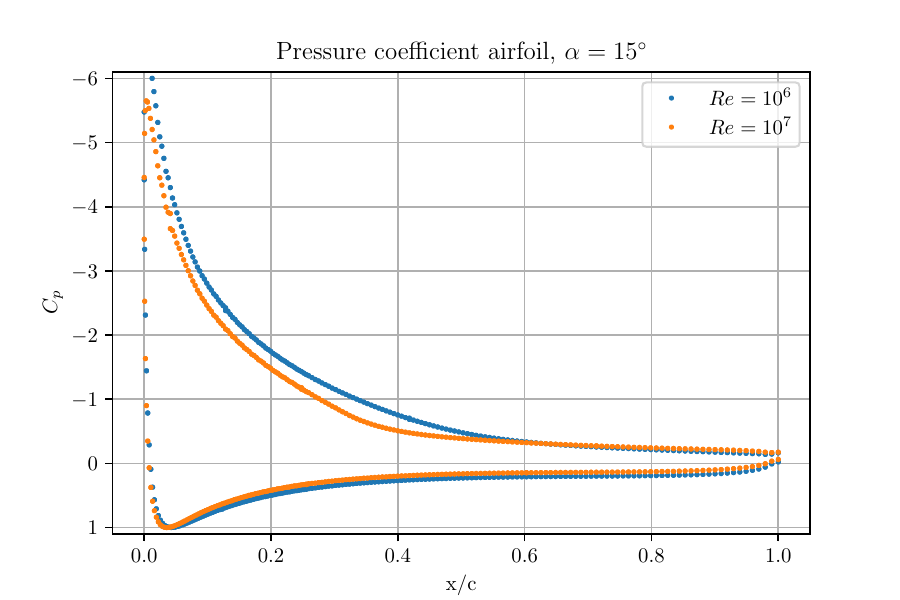}
      \end{minipage}\hfill
      \begin{minipage}{\mywidthTWO}
      \includegraphics[width=\textwidth]{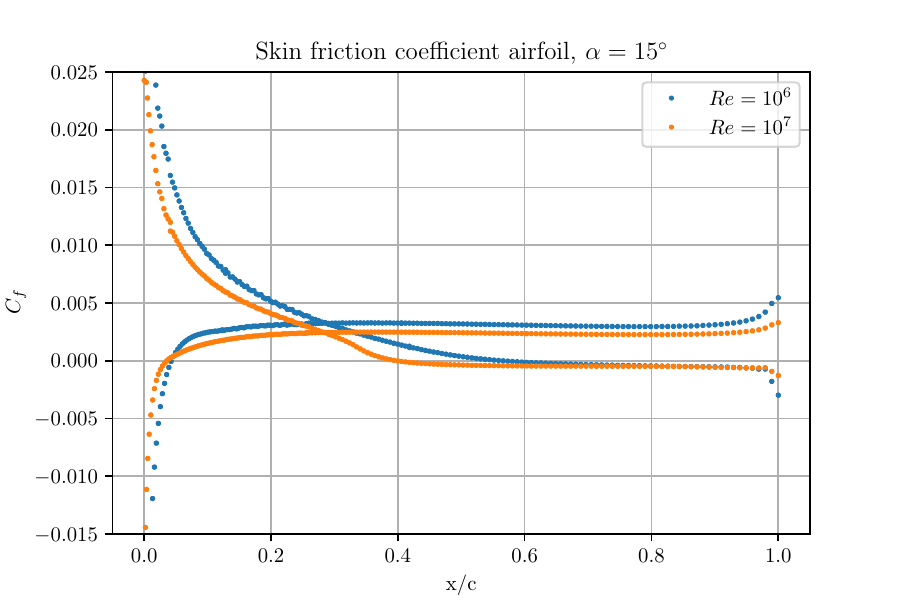}
      \end{minipage}
 \vspace{0.8em}
      \begin{minipage}{\mywidthTWO}
      \includegraphics[width=\textwidth]{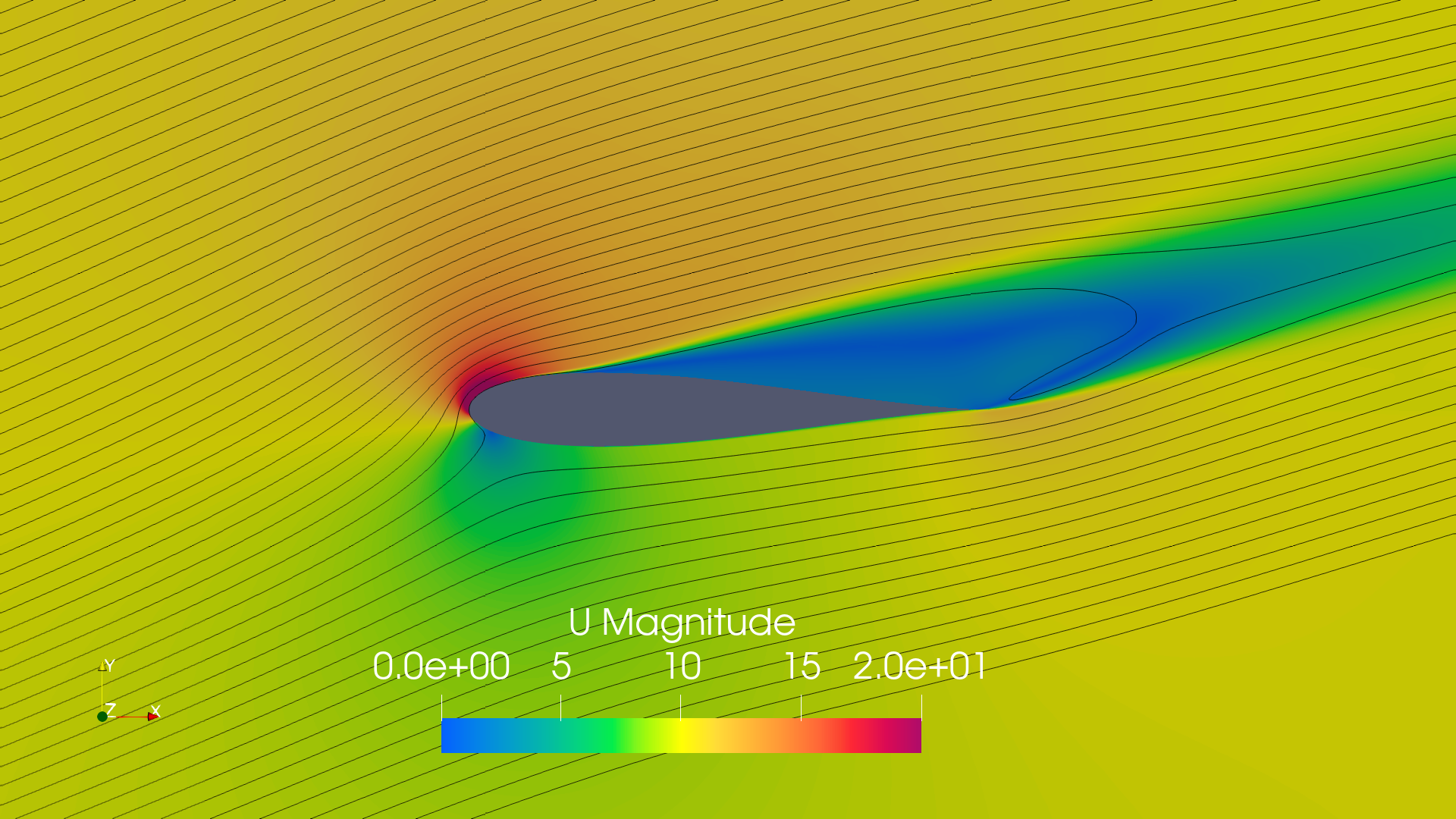}
      \end{minipage}\hfill
    \begin{minipage}{\mywidthTWO}
      \includegraphics[width=\textwidth]{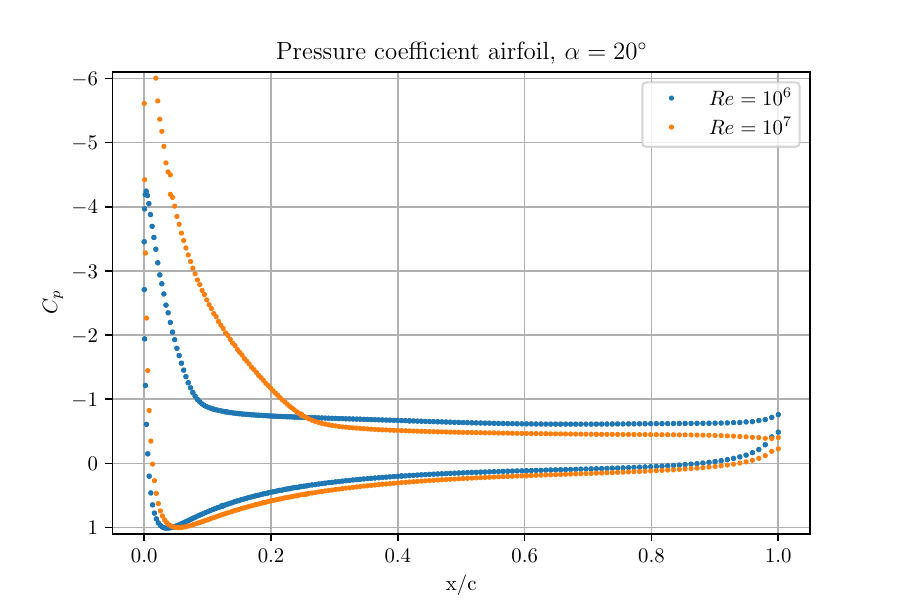}
      \end{minipage}\hfill
      \begin{minipage}{\mywidthTWO}
      \includegraphics[width=\textwidth]{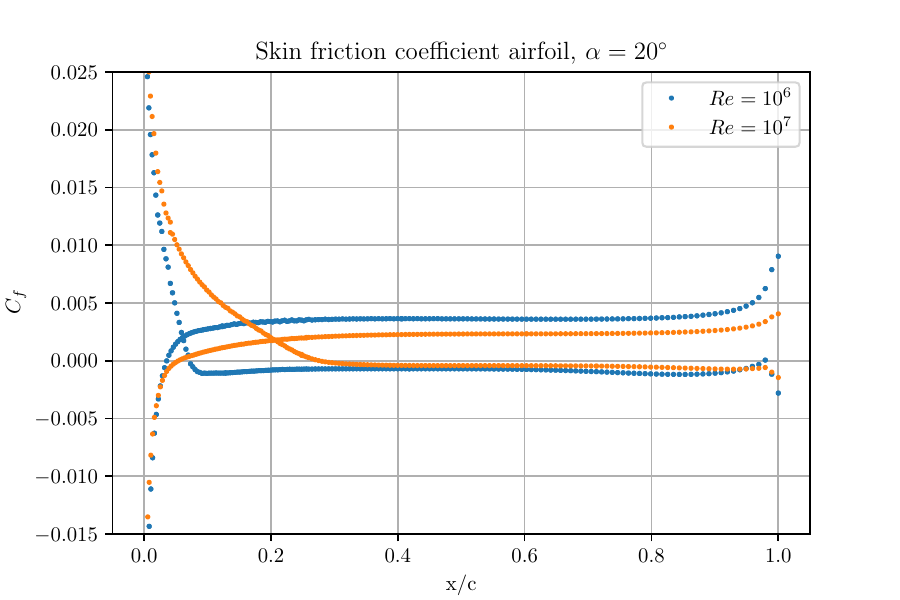}
      \end{minipage}
                  \caption{Simulation results for the airfoil, Each row, from top to bottom; $\alpha = 0^\circ$, $\alpha = 5^\circ$, $\alpha = 10^\circ$, $\alpha = 15^\circ$ and $\alpha = 20^\circ$ . Each column, from left to right; streamlines at $Re = 10^7$, Pressure coefficient and Skin friction coefficient.}
                        \label{fig:vk_billeder}
\end{figure}

The skin friction is normalized preserving the sign with respect to the x component (the dominant component of the velocity), enabling the later use of this to determine separation points. Here $ \tau $ is the shear stress acting on wall. Figures \ref{fig:ellispe_billeder} and \ref{fig:vk_billeder} show streamlines, pressure- and skin-friction distributions on the ellipse and airfoil respectively. 

\newpage

\subsection*{Force coefficients}
Forces are normalized according to equation \ref{eq:qinf}. 
Tables \ref{tab:forces_ellipse} and \ref{tab:forces_VK} show the lift and drag coefficients for the ellipse and airfoil respectively.

\vspace{1em}

\begin{minipage}{\mywidthdres}
\centering
\label{tab:liftandDrag}
\begin{tabular}{c|cc|cc}
\hline
{$\alpha$ [$^\circ$]} & \multicolumn{2}{c|}{Ellipse $Re = 10^6$} & \multicolumn{2}{c}{Ellipse $Re=10^7$} \\
\cline{2-5}
 & $C_d$ & $C_l$ & $C_d$ & $C_l$ \\
\hline
0 & 0.0238 & -0.0007 & 0.0163 & 0.0003 \\
1 & 0.0239 & 0.0799 & 0.0164 & 0.0761 \\
2 & 0.0242 & 0.1479 & 0.0166 & 0.1532 \\
5 & 0.0253 & 0.3769 & 0.0175 & 0.3895 \\
10 & 0.0278 & 0.7438 & 0.0194 & 0.7736 \\
15 & 0.0347 & 1.0316 & 0.0240 & 1.1005 \\
20 & 0.0662 & 1.0818 & 0.0365 & 1.3161 \\
\hline
\end{tabular}
\captionof{table}{Lift and Drag coefficients for the ellipse at different $\alpha$ and $Re$}
\label{tab:forces_ellipse}

\end{minipage}
\hfill
\begin{minipage}{\mywidthdres}

\centering
\label{tab:liftandDrag}
\begin{tabular}{c|cc|cc}
\hline
{$\alpha$ [$^\circ$]} & \multicolumn{2}{c|}{Airfoil $Re = 10^6$} & \multicolumn{2}{c}{Airfoil $Re=10^7$} \\
\cline{2-5}
 & $C_d$ & $C_l$ & $C_d$ & $C_l$ \\
\hline
0 & 0.0110 & 0.0000 & 0.0077 & 0.0000 \\
1 & 0.0111 & 0.1153 & 0.0080 & 0.1164 \\
2 & 0.0113 & 0.2304 & 0.0080 & 0.2351 \\
5 & 0.0133 & 0.5723 & 0.0093 & 0.5986 \\
10 & 0.0214 & 1.1081 & 0.0153 & 1.1506 \\
15 & 0.0481 & 1.4419 & 0.0533 & 1.2231 \\
20 & 0.2711 & 1.0832 & 0.1467 & 1.2043 \\
\hline
\end{tabular}
\captionof{table}{Lift and Drag coefficients for the airfoil at different $\alpha$ and $Re$}
\label{tab:forces_VK}
\end{minipage}

\subsection*{Stagnation and separation points}

\begin{figure}[h]
    \centering
    \includegraphics[width=0.65\linewidth]{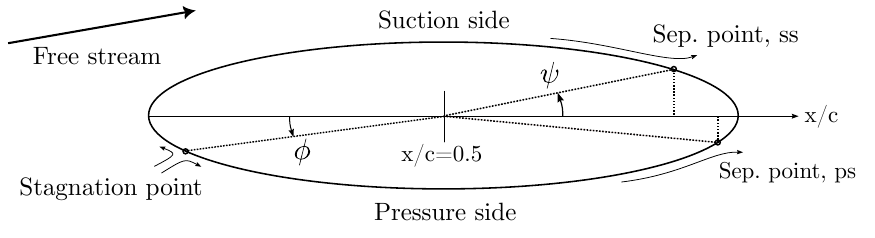}
    \caption{Sign convention for stagnation- and separation angles.}
    \label{fig:angle_sign_convention}
\end{figure}

Separation points are given in terms of $x/c$ as well as the parameter angle on the ellipse. See figure \ref{fig:angle_sign_convention} for sign convention of the angles. These angles are given in radians.

The stagnation points are identified as the point where the $C_p$ takes on its maximum value. Separation points are found where $C_f$ changes sign from positive to negative. 
Tabulated results for the stagnation points, and separation points and angles for the ellipse can be found in tables \ref{tab:EL_stag_sep_re106} and \ref{tab:EL_stag_sep_re107} for $Re = 10^6$ and $Re = 10^7$ respectively. Stagnation and separation points for the von-Kármán airfoil can be found in tables \ref{tab:VK_stag_sep_re106} and \ref{tab:EL_stag_sep_re107} for $Re = 10^6$ and $Re=10^7$ respectively.

\vspace{1em}
\begin{table}[h]
    \centering
    \label{tab:BCs}
    \begin{tabular}{@{}c|ccc|ccc@{}}
    \toprule
    $\alpha$ [$^\circ$]& stag. x/c & ss sep. x/c & ps sep. x/c & $\phi$ [rad] & ss-$\psi$ [rad] & ps-$\psi$ [rad] \\ \midrule
0 & 0.0000 & 0.9286 & 0.9287 & 0.0000 & 0.1491 & -0.1490 \\
1 & 0.0000 & 0.9249 & 0.9322 & 0.0000 & 0.1537 & -0.1444 \\
2 & 0.0005 & 0.9198 & 0.9362 & 0.0107 & 0.1603 & -0.1391 \\
5 & 0.0038 & 0.9026 & 0.9459 & 0.0312 & 0.1820 & -0.1262 \\
10 & 0.0132 & 0.8654 & 0.9565 & 0.0585 & 0.2294 & -0.1112 \\
15 & 0.0340 & 0.7945 & 0.9653 & 0.0969 & 0.3304 & -0.0980 \\
20 & 0.0526 & 0.6376 & 0.9721 & 0.1242 & 0.7179 & -0.0870 \\
    \bottomrule
    \end{tabular}
        \caption{Stagnation and separation points, ellipse $Re= 10^6$}
        \label{tab:EL_stag_sep_re106}
\end{table}

\begin{table}[H]
    \centering
    \label{tab:BCs}
    \begin{tabular}{@{}c|ccc|ccc@{}}
    \toprule
    $\alpha$ [$^\circ$]& stag. x/c & ss sep. x/c & ps sep. x/c & $\phi$ [rad] & ss-$\psi$ [rad] & ps-$\psi$ [rad] \\ \midrule
0 & 0.0000 & 0.9560 & 0.9560 & 0.0000 & 0.1119 & -0.1120 \\
1 & 0.0000 & 0.9535 & 0.9585 & 0.0000 & 0.1155 & -0.1082 \\
2 & 0.0005 & 0.9478 & 0.9607 & 0.0107 & 0.1236 & -0.1050 \\
5 & 0.0039 & 0.9403 & 0.9668 & 0.0312 & 0.1337 & -0.0956 \\
10 & 0.0170 & 0.9169 & 0.9738 & 0.0667 & 0.1640 & -0.0841 \\
15 & 0.0340 & 0.8756 & 0.9796 & 0.0969 & 0.2162 & -0.0736 \\
20 & 0.0575 & 0.7873 & 0.9845 & 0.1307 & 0.3420 & -0.0635 \\
    \bottomrule
    \end{tabular}
        \caption{Stagnation and separation points, ellipse $Re= 10^7$}
        \label{tab:EL_stag_sep_re107}
\end{table}

\begin{minipage}{\mywidthdres}

        \centering

        \label{tab:BCs}
        \begin{tabular}{@{}c|ccc@{}}
        \toprule
        $\alpha $ [$^\circ$] & stag. x/c & ss sep. x/c & ps sep. x/c \\ \midrule
    0 & 0.0000 & 1.0000 & 1.0000 \\
    1 & 0.0000 & 1.0000 & 1.0000 \\
    2 & 0.0000 & 1.0000 & 1.0000 \\
    5 & 0.0000 & 1.0000 & 1.0000 \\
    10 & 0.0223 & 1.0000 & 1.0000 \\
    15 & 0.0424 & 0.5646 & 1.0000 \\
    20 & 0.0355 & 0.0727 & 1.0000 \\
        \bottomrule
        \end{tabular}
        \captionof{table}{Stagnation and separation points, von-Kármán airfoil $Re=10^6$}
        \label{tab:VK_stag_sep_re106}
\end{minipage}
\hfill
\begin{minipage}{\mywidthdres}
        \centering
        \label{tab:BCs}
        \begin{tabular}{@{}c|ccc@{}}
        \toprule
        $\alpha$ [$^\circ$] & stag. x/c & ss sep. x/c & ps sep. x/c \\ \midrule
    0 & 0.0000 & 1.0000 & 1.0000 \\
    1 & 0.0000 & 1.0000 & 1.0000 \\
    2 & 0.0000 & 1.0000 & 1.0000 \\
    5 & 0.0000 & 1.0000 & 1.0000 \\
    10 & 0.0223 & 1.0000 & 1.0000 \\
    15 & 0.0355 & 0.3981 & 1.0000 \\
    20 & 0.0000 & 0.4848 & 0.4744 \\
        \bottomrule
        \end{tabular}
    \captionof{table}{Stagnation and separation points, von-Kármán airfoil $Re= 10^7$}
        \label{tab:VK_stag_sep_re107}
\end{minipage}

\vspace{1.5em}

The results generally follow expected trends. The skin friction coefficient is generally larger for smaller Reynolds number. A higher Reynolds number tends to slightly delay separation. The pressure distribution is affected very little by the Reynolds number, except in cases where there is a notable difference in the separation point.

\section{Discussion}

The presented work uses the RANS method for turbulence modeling as well as a steady state time scheme. It may be argued that separation phenomena that are highly transient are not well captured with this method. Generally, steady-state RANS will suppress phenomena such as von-Kármán vortex shedding, that may move the separation point in a transient manner.

The reason for this modeling strategy is twofold. Firstly, as the objective of this study is to provide reference data upon which extensions to potential flow models can be developed. Since these models are envisioned to be steady-state it is desirable to only have specially varying data. Secondly, steady state RANS is very computationally efficient compared to transient methods such as DES or LES. Due to the number of cases tested, computational efficiency was also a factor in the choice of overall modeling strategy.  



\section{Conclusion}

To further the research into potential flow models, that may handle separated flows, a systematic simulation study has been conducted. The study considers the flow about an ellipse and a von-Kármán airfoil. 

The simulations are verified by a mesh convergence study, analyzing the key results of interest: $C_l$, $C_d$ and separation points. For turbulence modeling, literature recommendations are used. Additionally, a turbulence sensitivity study has been conducted, showing little to no sensitivity within a reasonable range of the selected turbulent conditions, affirming the selection of these values. 

The simulations have been conducted at moderate Reynolds numbers of $Re = 10^6 $ and $Re = 10^7$, respectively. The geometries have been tested at a range of angles of attack, from 0$^\circ$ to 20$^\circ$. 

As for the results, $C_l$, $C_d$, $C_p$, $C_f$ as well as suction side and pressure side separation points are provided. Collectively these simulation results form a complete basis of comparison to validate many aspects of reduced-order flow models.



\section*{CRediT authorship contribution statement}

\paragraph{Christian Bak Winther} - Conceptualization, Investigation, Formal analysis, Validation, Writing – Original Draft, Writing – Review \& Editing, Visualization. 

\paragraph{Peter Ammundsen} - Investigation

\paragraph{Fynn Aschmoneit} - Conceptualization, Writing – Review \& Editing, Supervision

\section*{Acknowledgements}

The authors gratefully acknowledge the financial support of 
the Innovation Fond Denmark under grant no. 4365-00002B,
the Danish Maritime Fond under grant no. 2024-002 and
Orient Fonden.
The computational resources provided by ClAAUdia are also acknowledged.

\section*{Declaration of Competing Interest}
Authors declare no conflicts.

\bibliographystyle{plain}
\bibliography{references}

\end{document}